\documentclass[prd,onecolumn,notitlepage,nofootinbib]{revtex4-1}

\RequirePackage{fix-cm}

\RequirePackage{graphicx}

\usepackage[utf8]{inputenc}

\newcommand{\be}{\begin{equation}}
	\newcommand{\ee}{\end{equation}}

\newcommand{\bea}{\begin{eqnarray}}
	\newcommand{\eea}{\end{eqnarray}}

\usepackage{color}

\begin{document}

\title{Forecast and analysis of the cosmological redshift drift}

\author{Ruth Lazkoz$^{1}$}
\email{ruth.lazkoz@ehu.eus}
\author{Iker Leanizbarrutia$^{1}$}
\email{iker.leanizbarrutia@ehu.eus}
\author{Vincenzo Salzano$^{2}$}
\email{enzo.salzano@wmf.univ.szczecin.pl}

\affiliation{
	${}^1$ Department of Theoretical Physics, University of the Basque Country
	UPV/EHU, P.O. Box 644, 48080 Bilbao, Spain\\
	${}^2$ Institute of Physics, University of Szczecin, Wielkopolska 15, 70-451 Sczcecin, Poland}

\begin{abstract}
The cosmological redshift drift could lead to the next step in high-precision cosmic geometric observations, becoming a direct and irrefutable test for cosmic acceleration. In order to test the viability and possible properties of this effect, also called Sandage-Loeb (SL) test, we generate a model independent mock data set so as to compare its constraining power with that of the future mock data sets of Type Ia Supernovae (SNe) and Baryon Acoustic Oscillations (BAO). The performance of those data sets is analyzed by testing several cosmological models with the Markov chain Monte Carlo (MCMC) method, both independently and combining all data sets. Final results show that, in general, SL data sets allow for remarkable constraints on the matter density parameter today $\Omega_m$ on every tested model, showing also a great complementarity with SNe and BAO data regarding dark energy (DE) parameters.
\end{abstract}

\maketitle

\section{Introduction}
\label{intro}

Within the general relativity (GR) framework, no reliable explanation to the current acceleration of the universe exists which is simpler than
a $\Lambda$-term or cosmological constant \cite{Carroll:2000fy}. It  behaves as a fluid with negative pressure \cite{Barboza:2015rsa}, thus driving gravitational repulsion. This is, of course, also the kind of behaviour displayed by the plethora of other possible fluids so far proposed to try to accommodate data better than a cosmological constant. In broad terms, these settings that would cause the universe to accelerate are usually included in the so-called dark energy theories (see for reviews \cite{Li:2012dt,Kunz:2012aw,Copeland:2006wr,Bamba:2012cp,Mortonson:2013zfa}). There are, as well, other theoretical routes, with different levels of complexity (not necessarily unrelated \cite{Battye:2015hza}), which venture to modify GR.

This background expansion of the universe can be measured with a lot of different probes: luminosity distances from Type Ia Supernovae (SNe) \cite{Riess:1998cb,Perlmutter:1998np,Suzuki:2011hu,Beutler:2011hx}; the acoustic peaks in the Cosmic Microwave Background (CMB) \cite{Ade:2015xua,cmb2}; and their counterpart imprinted in clustered matter, i.e. Baryon Acoustic Oscillations (BAO) \cite{sdss1,sdss2,bao2,bao3,Blake2012}; or through the matter power spectrum obtained from weak lensing \cite{Weinberg:2012es,Bartelmann:1999yn}. Usually, a time integral along redshift connects those data with the expansion rate/history of the universe, the Hubble parameter $H(z)$, and are enough to constrain quite satisfactorily the geometry and energy content of the universe.

On the other hand, one expects that the expansion of the universe will make the redshift of a given astrophysical object exhibit a drift over time,  which should in principle be amenable to giving an accurate description of that very same expansion once an underlying model is chosen. While looking for a possible temporal variation of the redshift of extra-galactic sources, Sandage came  in 1962 \cite{Sandage:1962} to the conclusion that it should indeed occur. But, alas, the limited technological resources on deck  at that epoch, lead to the inference that a measurement time interval of the order of $10^7$ years would be required for a signal detection. When new spectroscopic techniques became available to astrophysicists, Loeb paid a new visit to the concept \cite{Loeb:1998bu} in 1998, and concluded that the new technology would allow a reduction in the observation time interval to a few decades. This cosmological redshift drift measurement, also called Sandage-Loeb (SL) test, would then provide a direct proof of the accelerated expansion of the universe. In fact, this temporal variation is directly related to the expansion rate at the source redshift, being thus a direct measurement of the Hubble function.

The last results of the \textit{Planck} survey \cite{Ade:2015xua}, have made us enter an ultra high precision cosmology era; and other future surveys are scheduled which should further improve the accuracy of cosmological measurements, as for example \textit{Euclid} \cite{Laureijs:2011gra}, \textit{Wide-Field Infrared Survey Telescope (W-First)} \cite{Spergel:2013tha} or Square Kilometer Array (SKA) \cite{SKA2015a}. 

Thus, in the near future, available resources will allow us to start thinking about the next level of cosmological observational data, to which the cosmological redshift drift will contribute, complementing the previously cited surveys. However, even with future precision radio telescopes, the measurement of the SL effect represents a difficult enterprise \cite{Killedar:2009xw}, as it demands several years of observation (usually some decades) to register enough signal-to-noise ratio so as to yield a possible reliable detection of the cosmological redshift drift signal. Best candidate objects for a feasible detection of this faint signal are good Hubble flow tracers as far as possible \cite{Martinelli:2012vq}. As put forward by Loeb \cite{Loeb:1998bu}, an auspicious target would be the Lyman-$\alpha$ forest measurements of distant quasars (QSO). With spectroscopic techniques to be operational in the near future, like CODEX (COsmic Dynamics and EXo-earth experiment) experiment \cite{Liske:2008ph}, proposed for the European-Extremely Large Telescope (E-ELT), or radio telescopes as SKA \cite{Klockner:2015rqa}, these observations will grant access to direct measurements of the Hubble parameter up to redshift $ \sim 5$, a so far not yet observed redshift range. Thus, the SL test will open a new ``cosmological window''.

Due to the near future possibilities to measure the cosmological redshift drift, this type of observations has recently drawn some attention. The reconstruction of the theoretical SL signal that different cosmological models would produce has been explored quite extensively \cite{Zhang:2013fqa,Vielzeuf:2012zd}. It comes out that the range and variety of the different cosmological redshift drift signals created by various models is remarkable: from those created by different proposals for dark energy's equation of state or modified gravity \cite{Corasaniti:2007bg,Moraes:2011vq}, to the ones created by backreaction in an inhomogeneous universe without the presence of dark energy \cite{Koksbang:2015ctu}; from the peculiar signals for Lema\^itre-Tolman-Bondi models \cite{Yoo:2010hi,Araujo:2010ir}, to even a null signal \cite{Melia:2016bnb} for the $R_{h}=c t$ Universe, or other several exotic scenarios \cite{Denkiewicz:2014kna,Zhang:2010im,Banerjee:2015ala,Mishra:2012vi,Balcerzak:2012bv,Balcerzak:2013kha}. SL signals have been used as an hypothetical geometric cosmic discriminant \cite{Guo:2015gpa,Geng:2014hoa,Geng:2014ypa} to show the corresponding improvement in the constraints that can be achieved due to the degeneracy breaking (around $20\%$ of improvements for dark energy parameters and even $65\%$ for matter density). SL mock data sets have been applied with similar results as cosmic observational discriminators to test other various models, like interactive dark energy models \cite{Geng:2015ara,Zhang:2013zyn}, modified gravity \cite{Geng:2015hen,Li:2013oba}, and other exotic cosmologies \cite{Zhang:2007zga,Zhu:2015pta}. Their power to differentiate models has been exploited also in the context of the model-independent approach of cosmography \cite{Zhang:2014bwt,Martins:2016bbi}. Besides, some new approaches \cite{Kim:2014uha} can lead to ambitious ideas, such as real-time cosmology \cite{Quercellini:2010zr}.

We stress again the fact that the measurement of the cosmological redshift drift is not an easy pursuit, and requires quite a lot of planning due to the large observation time interval of the survey. Thus, foreseeing the contribution and behaviour of this type of measurements is important, and we precisely carry out here a quite thorough forecast analysis of cosmological redshift drift constraints on various cosmological models. The analysis includes a comparison between the proposed SL data with other future planned surveys, generating mock data based on the given specifications. Furthermore, unlike previous works, all mock data sets are generated in a fully model independent way, with no fiducial cosmological model chosen to generate the points. In Sec.~\ref{sec:SL_Theory} we introduce the mathematical formalism of the cosmological redshift drift, and then we give the details of the  mock data sets we use for our predictions. We find it convenient to produce a SL data set, but also auxiliary SNe and BAO data. In Sec.~\ref{sec:MCMC}, we explain our MCMC procedure which will eventually constrain the cosmological models we have chosen as reference. Finally, in Sec.~\ref{sec:Results}, we present and discuss the outcomes of that statistical analysis, and then summarize and outline the main conclusions.

\section{Cosmological Redshift drift}
\label{sec:SL_Theory}

A preliminary straightforward calculation introduces the main observable quantity we are going to focus on, i.e. the cosmological redshift drift, (see for example \cite{Liske:2008ph} or \cite{Corasaniti:2007bg}). In an homogeneous and isotropic universe with a Friedmann-Robertson-Walker metric, let us consider a source at rest emitting electromagnetic waves isotropically, without any (significant) peculiar velocity. Thus, the comoving distance between the source and an observer can be considered fixed. If the source emits electromagnetic waves during time $(t_e, t_e+\delta t_e)$, and they are detected by the observer in the interval $(t_o, t_o+ \delta t_o)$, where $t_e$ is the emission time and $t_o$ is the time they reach the observer, then the following relation is satisfied:
\begin{equation}
\int_{t_e}^{t_o} \frac{dt}{a(t)} = \int_{t_e+\delta t_e}^{t_o+\delta t_o} \frac{dt}{a(t)}  \; ,
\label{r.1}
\end{equation}
provided the universe through which the waves travel is a spatially flat Friedmann-Robertson-Walker spacetime. If the time intervals are small $(\delta t_e, \delta t_o \ll t_e, t_o)$, the above expression leads to the well known redshift relation between the emitted and the observed radiation:
\be
\frac{\delta t_e}{a(t_{e})} = \frac{\delta t_o}{a(t_o)} \quad \Rightarrow \quad \frac{\lambda_o}{\lambda_e} = \frac{a(t_o)}{a(t_e)} = 1+z_e(t_o) \, ,
\label{r.2}
\ee
where $z_{e}(t)$ is the redshift of the source as measured at a certain observation time $t_o$. Other waves can be emitted by the source $\delta t_e$ time later, specifically, at time $t_e + \delta t_e$, and they will be observed at $t_o + \delta t_o$. In the case of these waves, it is straightforward to modify Eq.~\ref{r.2} regarding the new time periods and redshift. Thus, the observer can measure the difference between the redshifts observed at $t_o$ and $t_o +\delta_o$:
\be
\Delta z_e=z_e(t_o+\delta t_o)-z_e(t_o)=\frac{a(t_o+\delta t_o)}{a(t_e+\delta t_e)}-\frac{a(t_o)}{a(t_e)} \; .
\ee
Within the $\delta t / t \ll 1$ approximation, the first ratio can be expanded to linear order:
\be
\frac{a(t_o+\delta t_o)}{a(t_e+\delta t_e)} \simeq \frac{a(t_o)}{a(t_e)} + \frac{\dot{a}(t_o) \delta t_o}{a(t_e)} - \frac{a(t_o) \dot{a}(t_e) \delta t_e}{a(t_e)^2} \; .
\label{1stOrder}
\ee
Inserting Eq. \ref{r.2} into the first order expansion in Eq. \ref{1stOrder}, an approximated expression for the redshift variation can be obtained:
\be
\Delta z_e \simeq \left[ \frac{\dot{a}(t_o)-\dot{a}(t_e)}{a(t_e)}\right] \delta t_o \; .
\ee
Under the assumption that the observation time is today, we normalize by letting the corresponding scale factor satisfy $a(t_o)=1$; and then, using both the Friedmann equation and the known redshift relation Eq.~\ref{r.2}, we can rewrite the above expression in terms of the Hubble parameter $H(z)=\dot{a}(z)/a(z)$:
\be
\Delta z_e = \delta t_o \left[ H_0(1+z_e) - H(z_e) \right] \; ,
\ee
with $H_0=H(z_0)$ being the Hubble constant today. This redshift variation can be expressed as a spectroscopic velocity shift $\Delta v \equiv c \Delta z_e /(1+z_e)$, and using the dimensionless Hubble parameter $E(z)=H(z)/H_0$, we get the final expression
\be
\Delta v = c H_0 \delta t_o \left[ 1 - \frac{E(z_e)}{1+z_e}  \right] \; .
\label{v-shift}
\ee

\subsection{Sandage-Loeb mock data set}

\begin{figure*}
	\includegraphics[width=0.45\textwidth]{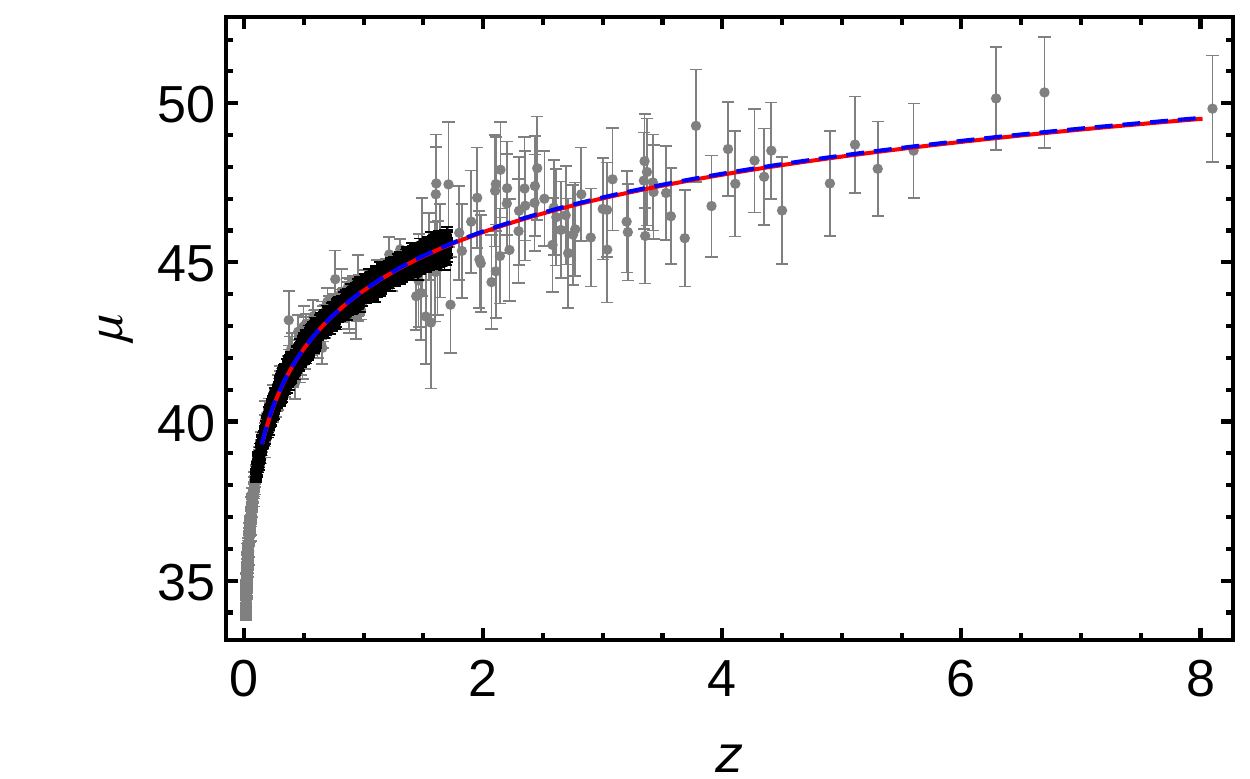}
	\includegraphics[width=0.45\textwidth]{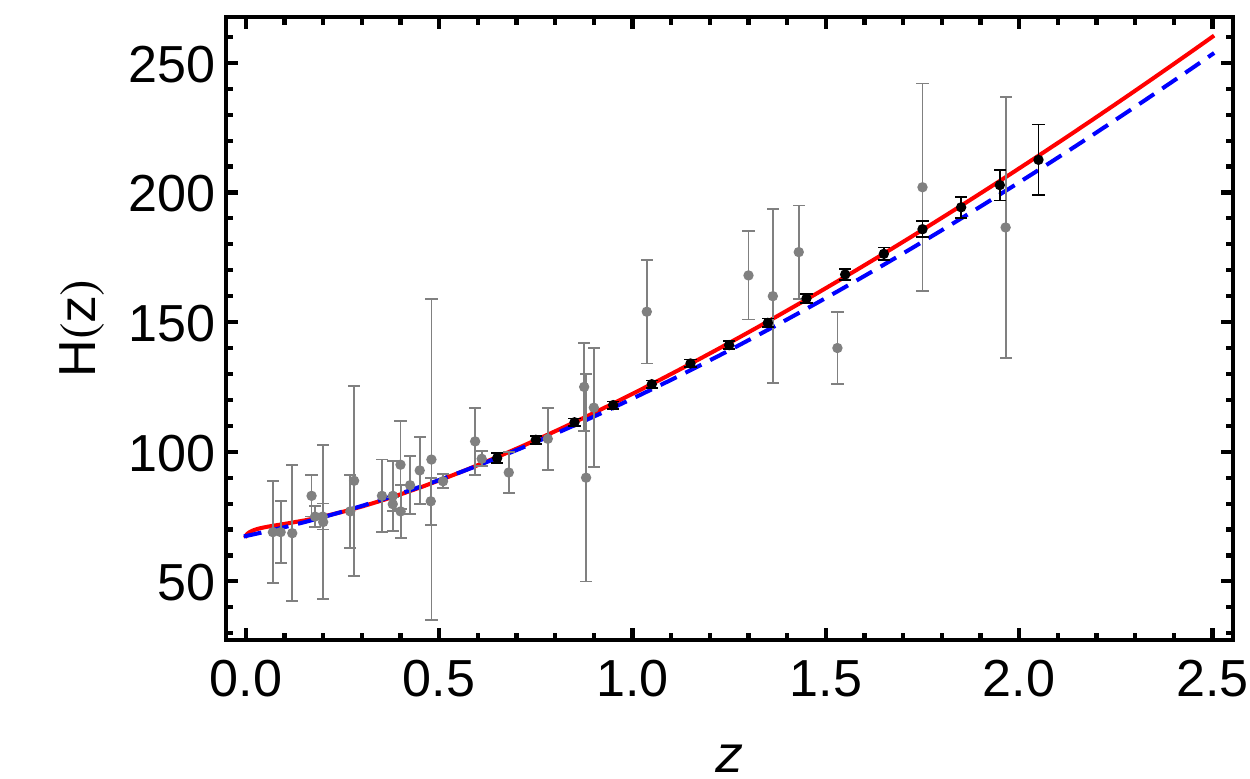}\\
	\includegraphics[width=0.45\textwidth]{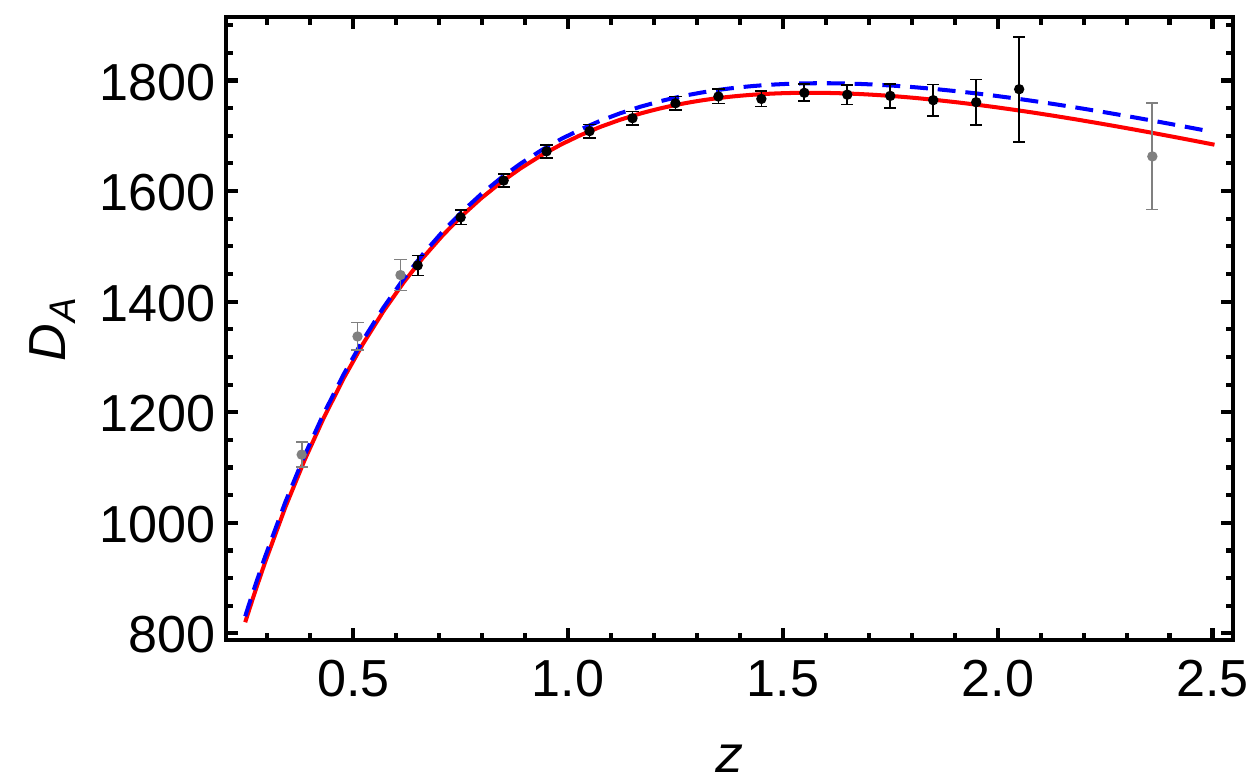}
	\includegraphics[width=0.45\textwidth]{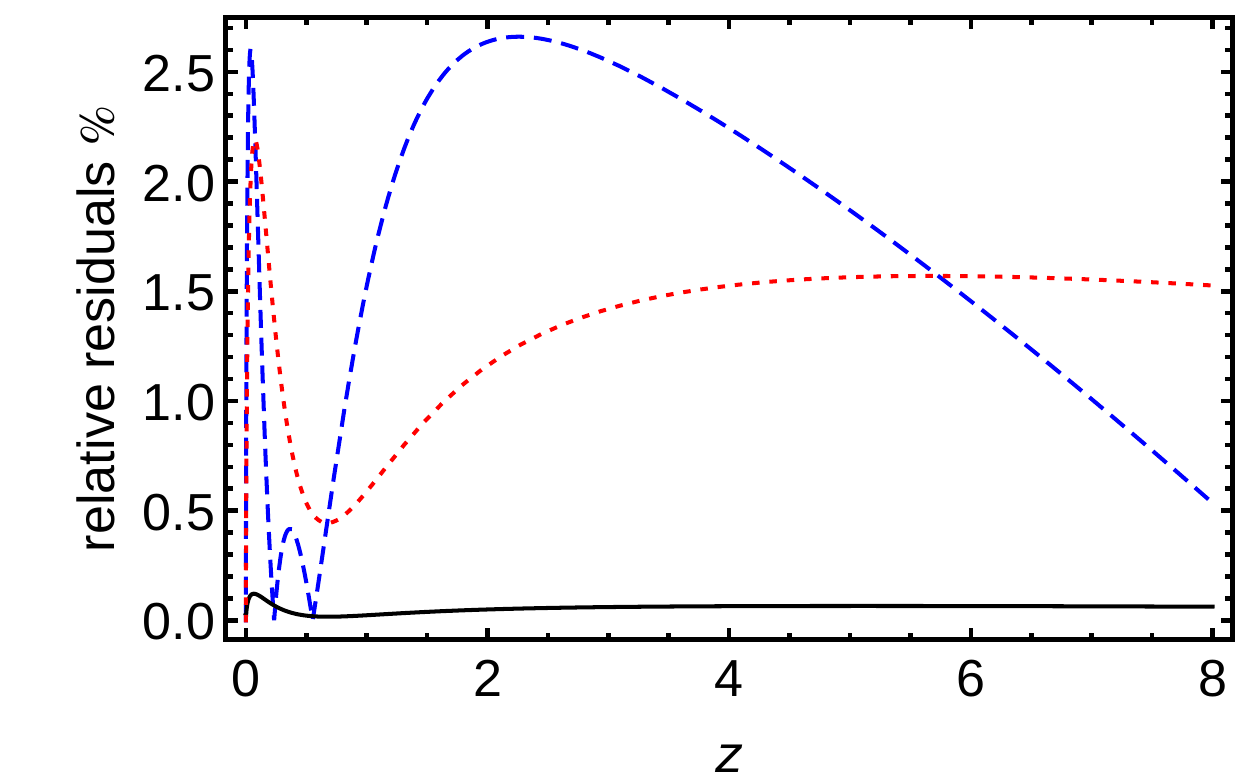}
	\caption{Top left panel: Comparison between the selected phenomenological function $F_{fit}(z;b,c,d,e)$ given in Eq.~(\ref{f-fit}) (solid red) with the \textit{Planck} $\Lambda$CDM (dashed blue) described in the text. Grey dots and bars are distance modulus values and related errors for SNe Union and Mayflower GRBs samples, and black ones are for our generated SNe mock data. Top right panel: comparison between the $H(z)$ function derived from Eq.~(\ref{f-fit}) (solid red) with that corresponding to the \textit{Planck} $\Lambda$CDM (dashed blue) described in the text. Grey dots and bars are expansion rate values and related errors from cosmic chronometers and black ones our generated mock data. Bottom left panel: comparison of the $D_A(z)$ function derived from Eq.~(\ref{f-fit}) (solid red) with that corresponding to the \textit{Planck} $\Lambda$CDM (dashed blue) case described in the text. Grey dots and bars are angular diameter distances values and related errors from BOSS and SDSS, and black ones our generated mock data. Bottom right panel: relative residuals between our model and the \textit{Planck} $\Lambda$CDM for the Hubble function (dashed blue), the distance modulus (solid black) and the angular diameter distance (dotted red).}
	\label{fig:comparison}       
\end{figure*}

In order to generate our SL observational mock data set in a fully model independent manner, we try to derive a Hubble function from a phenomenological distance modulus, in a fashion similar to \cite{Padmanabhan:2002vv}. We propose this observable because it is well measured by Type Ia Supernovae (SNe) and can be extended to high redshifts, even if with lower precision, by Gamma Ray Bursts (GRBs, Mayflower sample) \cite{Liu:2014vda}. We model this phenomenological distance modulus as
\be
\mu_{fit}(z) = a + 5 \log_{10} \left[ F_{fit}(z;b,c,d,e) \right] \; ,
\label{mu-fit}
\ee
where $F_{fit}$ is an {\it ad hoc} proposed function (among many) mimicking the luminosity distance. This phenomenological function is then fitted using the SNe data set Union 2.1 \cite{Suzuki:2011hu} for the low-redshift regime, and the GRBs sample calibrated by the Pad\'{e} Method \cite{Liu:2014vda} for the high-redshift one. Once $\mu_{fit}$ is fitted, other observational quantities relevant to our work can be easily obtained. For instance, the Hubble function can be derived recalling the relation:
\be
\mu(z) = 5 \log_{10} d_L(z) + \mu_{0} \; ,
\label{mu}
\ee
where, in the spatially flat universe we are considering, the dimensionless luminosity distance $d_L$ is defined as
\be
d_L(z) = (1+z) \int_0^z \frac{dz'}{E(z')} \; ,
\label{lum}
\ee
and $\mu_0$ stores all the information related to the constants involved such as the speed of light $c$, the Hubble constant $H_0$, and the SNe absolute magnitude. By comparing both distance moduli, $\mu$ from Eq.~(\ref{mu}) and $\mu_{fit}$ from Eq.~(\ref{mu-fit}), one can realize that the dimensionless luminosity distance $d_L(z)$ is equivalent to the function $F_{fit}$. Thus, the dimensionless Hubble function is
\be
E_{fit}(z)= \left( \frac{d}{dz} \frac{F_{fit}(z;b,c,d,e)}{(1+z)} \right)^{-1} \; .
\label{E-fit}
\ee
Once such phenomenological dimensionless Hubble parameter $E_{fit}(z)$ is obtained, we can ``mimick'' all the cosmological probes we need for our analysis, as they are all related to it. In this way, we can create cosmological-model-independent mock data sets, where the only intrinsic information we are using for $E_{fit}$ is that it has to be able to fit present data (in this case, SNe and GRBs). Of course, some arbitrariness lies behind the choice of the phenomenological function $F_{fit}$; we have tried to use the most general type of functions possible, and we have selected the best one based on a simple best-fitting (minimum $\chi^2$) criterium. The best performing function we have found is
\be
F_{fit}(z;b,c,d,e) =  \frac{z (1 + b \log[1 + z]^{d})}{(1 + c \log[1 + z]^{e})} \; ,
\label{f-fit}
\ee
where the values for the parameters are shown in Table \ref{table:f-fit}. It can be seen in Fig. \ref{fig:comparison}, in the top left panel, that this function fits the distance modulus points of the  Union 2.1 \cite{Suzuki:2011hu} and Mayflower \cite{Liu:2014vda} data sets as much satisfactorily as a $\Lambda$CDM with \textit{Planck} values, $\Omega_{m}=0.3121$ (sixth column of Table~4 in \cite{Ade:2015xua}). In the top right panel, we also compare the expansion rate function $H$ which can be derived from Eq.~(\ref{f-fit}) with the same \textit{Planck} $\Lambda$CDM and with data from cosmic chronometers \cite{Moresco:2016mzx}. In the bottom left panel, the comparison between angular diameter distance derived from Eq.~(\ref{f-fit}) and the same \textit{Planck} $\Lambda$CDM is done, with the data coming as comoving angular diameter distance from galaxy clustering (BAO+FS column of Table 7 in \cite{Alam:2016hwk}) and physical angular diameter distance coming from quasar cross-correlation (Eq.~(21) in \cite{bao3}). Finally, in the bottom right panel, we can also appreciate that the difference between our model and the \textit{Planck} $\Lambda$CDM is minimal for the case of the distance modulus $(\sim 0.1\%)$, and small for both the Hubble function $(\sim 2.5\%)$ and the angular diameter distance  $(\sim 2\%)$, all over the redshift range we cover with our mock data in our analysis.

\begin{table}
\centering	
	\caption{Parameter values of $F_{fit}$.}
	\label{table:f-fit}       
	\begin{tabular}{lll}
		\hline\noalign{\smallskip}
		  & Estimate & Standard Error  \\
		\noalign{\smallskip}\hline\noalign{\smallskip}
		a & 43.2025 & 0.146659 \\
		b & 2.29876 & 1.60875 \\
		c & 0.92048 & 0.969826 \\
		d & 1.05317 & 0.62311 \\
		e & 0.814751 & 0.922533 \\
		\noalign{\smallskip}\hline
	\end{tabular}
\end{table}

Once we have our $E_{fit}(z)$, we only need to specify a fiducial value for the Hubble constant to insert in Eq.~(\ref{v-shift}), whose effect is only the rescaling of the velocity shift value. We fix the value of $H_0= 67.51 \, km/s/Mpc$ from the TT,TE,EE + lowP + lensing baseline model of \textit{Planck} \cite{Ade:2015xua}. Then, for what concerns SL data, the points lie in the redshift range $2 < z < 5$, randomly distributed within the following bins: $2<z<3$ (13 points), $3<z<3.5$ (7 points), $3.5<z<4$ (4 points), $4<z<4.5$ (3 points) and $4.5<z<5$ (3 points). This way, we try to mimick the reduction of the number of data points while increasing the redshift as in \cite{Risaliti:2015zla}.

According to Monte Carlo simulations carried out to eventually mimick results from CODEX \cite{Liske:2008ph,CODEX2010a}, the standard deviation on the measured spectroscopic velocity shift $\Delta v$ can be estimated as
\be
\sigma_{\Delta v} = 1.35 \frac{2370}{S/N} \sqrt{ \frac{30}{N_{QSO}}} \left( \frac{5}{1+z_{QSO}} \right)^x {\rm cm \, s^{-1}} \; ,
\label{error-v-shift}
\ee
where $x$ is $1.7$ for $z\leq 4$, and $0.9$  beyond that redshift, $S/N$ is spectral signal to noise ratio of Ly-$\alpha$, $N_{QSO}$ is the number of observed quasars and $z_{QSO}$ their redshift. The error for the mock data is given by assuming a fix number of integration time hours which yields a value of $S/N= 3000$ for the signal-to-noise ratio and $N_{QSO}=30$ for the number of quasars observed \cite{Martinelli:2012vq}. We also introduce some noise to disperse the data points around the fiducial value derived from $E_{fit}$, using a Gaussian distribution centered on such values, and with a standard deviation corresponding to the expected error on the SL observation, $\sigma_{\Delta v}$, obtained by error propagation from the fitted parameters of the selected function Eq.~(\ref{f-fit}).\\

\begin{figure*}
	\includegraphics[width=0.45\textwidth]{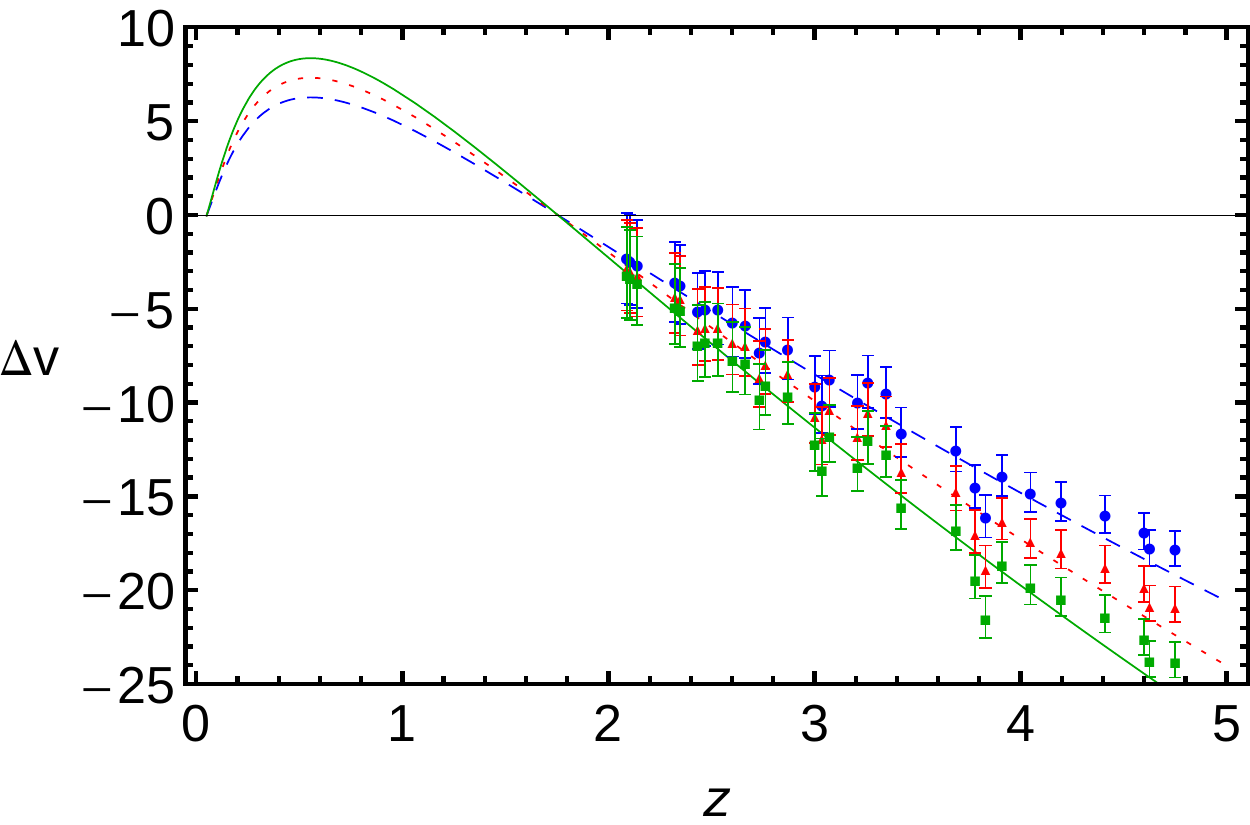}
	\includegraphics[width=0.45\textwidth]{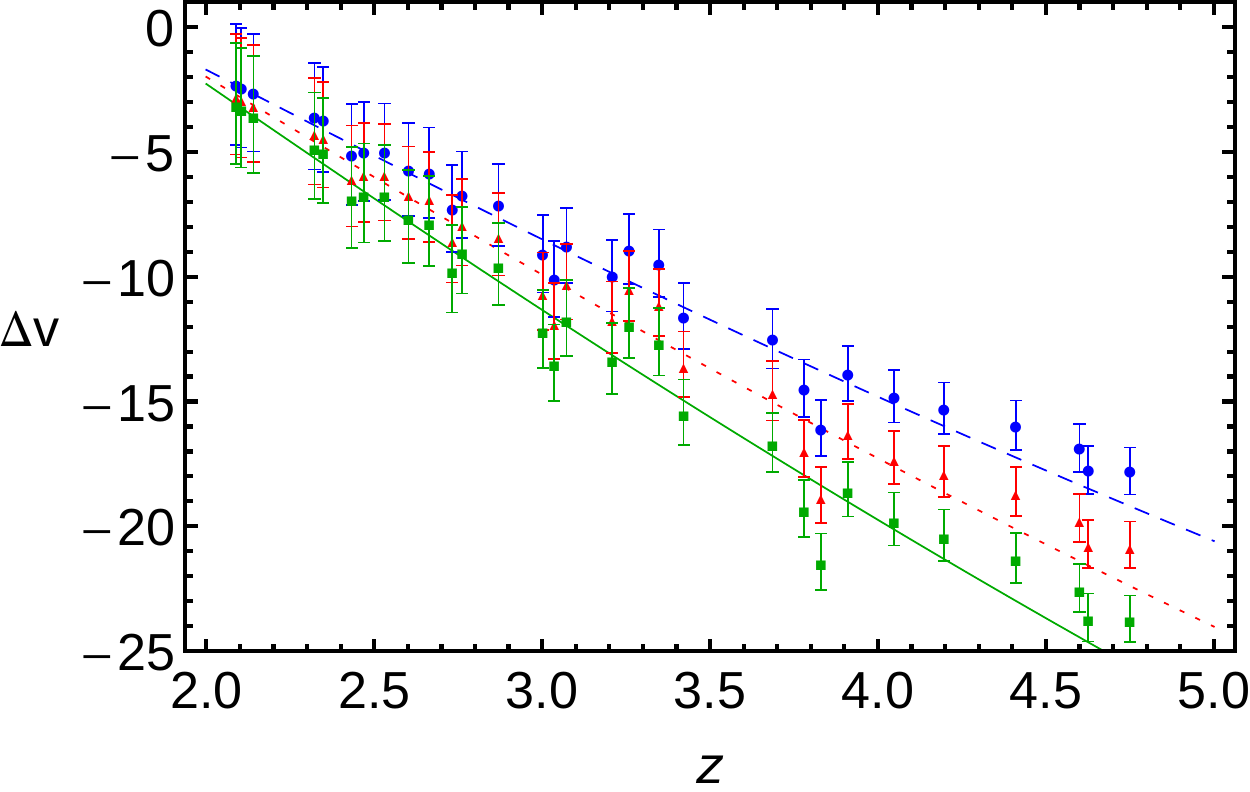}
	\caption{Datasets for SL test based on $F_{fit}(z;b,c,d,e)$ for different observation periods: blue (circle and dashed line) for 24 years, red (triangle and dotted line) for 28 years and green (square and solid line) for 32 years.}
	\label{fig:SL-data sets}       
\end{figure*}

Note that the magnitude of the observed cosmological redshift drift is proportional to the observation period, although the error does not dependend on it. Thus, once a data set for some given observational time $\Delta t_A$ is created, any new mock data set with different observation period $\Delta t_B$ can be easily calculated by
\be
\Delta v_B = \frac{\Delta t_B}{\Delta t_A} \Delta v_A \; .
\ee
We use three observation periods of 24, 28 and 32 years, which are the most illustrative among the data sets tested. The resulting data sets for SL test can be seen on Fig.~\ref{fig:SL-data sets}.

\subsection{Auxiliary Mock Datasets}

We include additional future mock data sets alongside the cosmological redshift drift data set so as to constrain models better. Basically, the reason why we introduce in the picture these other probes is our interest on studying and quantifying the relative performance of SL with respect to more standard and used probes, and our aim of finding out whether the cosmological redshift drift data have some degree of complementarity with them, thus providing eventual tighter constraints. These auxiliary mock data sets are created from the same model independent function Eq.~(\ref{mu-fit}).

\subsubsection{\textit{W-First} SNe}

The first mock data set we produce is a SNe catalogue  based on the \textit{W-First} forecast \cite{Spergel:2013tha}, which includes $2725$ SNe randomly picked in redshift bins of $\delta z = 0.1$ spread through a redshift range of $0.1 < z < 1.5$ according to the distribution given by \cite{Spergel:2013tha}.

Given that in the SNe case what one measures is the distance modulus, we can make direct use of the fitted function Eq.~(\ref{mu-fit}) to generate the mock data points. As in the SL case, we also introduce some Gaussian noise to disperse the data points around the mean value.

To create the error bars for this catalogue and the dispersion for the Gaussian noise, we use the information given in \cite{Spergel:2013tha}. The statistical errors they account for are the following: the photometric measurement error, $\sigma_{meas}=0.8$; the intrinsic luminosity dispersion $\sigma_{int}=0.08$; and the gravitational lensing magnification $\sigma_{lens}=0.07$. Besides, they assume a systematic error $\sigma_{sys}=0.01 (1+z)/1.8$. Thus the total error per SNe is:
\be
\sigma_{tot}=\sqrt{ \sigma_{stat}^2 + N_{\rm SN }\sigma_{sys}^2 } \; ,
\label{SN-error}
\ee
where $\sigma_{stat}=\sqrt{\sigma_{meas}^2+\sigma_{int}^2+\sigma_{lens}^2}$ and $N_{\rm SN}$ is the number of SNe in the bin. The data set generated for the \textit{W-First} SNe survey is shown on Fig.~\ref{fig:comparison}.

\subsubsection{\textit{Euclid} BAO}

The second data set we consider is BAO. We choose the future \textit{Euclid} survey \cite{Laureijs:2011gra} as the experiment to reproduce. The two quantities we consider are the angular diameter distance
\be
D_A(z) = \frac{c}{1+z} \int_0^z \frac{dz'}{H(z')}
\ee
normalized by the sound horizon $D_A(z)/r_s$, and the Hubble parameter times the sound horizon, $H(z) \, r_s$, where the value of $r_s=144.71$ $Mpc$, consistent with the previous $H_{0}$, is used according to \cite{Ade:2015xua}.

Both the angular diameter distance and the Hubble parameter are reconstructed using again the function Eq.~(\ref{mu-fit}). We have already discussed that the Hubble parameter can be inferred as in Eq.~(\ref{E-fit}), once a value for $H_{0}$ is decided. Instead, in order to derive the angular diameter distance from the same Eq.~(\ref{mu-fit}), we use its definition and its relation with the luminosity distance $(1+z)^2 D_A=D_L$, thus to obtain
\be
D_A(z) = \frac{c}{H_0} \frac{F_{fit}(z;b,c,d,e)}{(1+z)^2} \; .
\label{fit_DA}
\ee
The redshift values of the data set are taken from \cite{Font-Ribera:2013rwa}, and they specifically are the central redshfits of 15 bins with $\delta z =0.1$ width, spread from $z=0.5$ to $z=2.1$. The error in each redshift value for both $D_A$ and $H_0$ is build from the percentage error given also in \cite{Font-Ribera:2013rwa}. Finally, we introduce some Gaussian noise using the error from each bin as dispersion when generating the points $D_A(z)/r_s$ and $H(z) \, r_s$. The resulting data sets can be seen at Fig.~\ref{fig:comparison} before normalizing the observables by the comoving sound horizon $r_s$.

\section{Testing Models}
\label{sec:MCMC}

Within the Bayesian framework, we wish to find out how SL constrains the probability distribution function of some cosmological parameters. For that purpose, we need the posterior distribution, or equivalently the likelihood, which can be straightforwardly computed with MCMC sampling while minimizing the $\chi^2$ function. The knowledge of the posterior probability gives a better and more complete information about the parameters, including the full correlation among them.

Thus, once we have the mock data sets, we build the $\chi^2$ function for each observable and, once all contributions are summed up, we minimize the total $\chi^2$ in order to perform our statistical analysis. The $\chi^2$ contribution for the spectroscopic velocity shift is simply
\be
\chi^2_{SL} = \sum_i \left( \frac{ \Delta v^{theo}_i - \Delta v_i ^{mock}}{\sigma_{\Delta v_i}}  \right)^2  \; ,
\label{chi-v-shift}
\ee
where $\Delta v^{theo}_{i} = \Delta v(z_i)$ follows from Eq.~(\ref{v-shift}), while errors $\sigma_{\Delta v_i}$ are given by Eq.~(\ref{error-v-shift}). The errors are arranged into a diagonal covariance matrix. Depending on whether the SL surveys will use overlapping redshift bins or not, the error could be more realistically given by a non-diagonal covariance matrix. As we lack such information, we adopt the optimistic diagonal covariance matrix assumption, always keeping in mind that it could lead to a general underestimation of the global errors on the cosmological parameters. The period of observation $\Delta t_o$, that will be specified below, changes depending on the mock SL survey tested. In the case of the $\chi^2$ contribution of SNe, the $\chi^2$ reads:
\be
\chi^2_{\rm SN}= \sum_i \frac{(\mu(z_i) - \mu^{mock}_i)^2} {\sigma_{\mu_i}^2}\ ,
\label{chi-SN1}
\ee
where the error is given by Eq.~(\ref{SN-error}). We can marginalize $\chi^2$ over the parameter $\mu_0$ by expanding the $\chi^2$ in Eq.~(\ref{chi-SN1}) with respect to $\mu_0$ as
\begin{equation}
\chi^2_{\rm SN}= A - 2 \mu_0 B  + \mu_0^2 C\ ,
\label{SN6}
\end{equation}
where
\bea
A &=&\sum_i \frac{(\tilde{\mu}(z_i) - \mu^{mock}_i)^2}{\sigma_{\mu_i}^2}\, , \\
B &=&\sum_i \frac{ \tilde{\mu}(z_i) - \mu^{mock}_i }{\sigma_{\mu_i}^2}\, , \nonumber \\
C &=&\sum_i \frac{1}{\sigma_{\mu_i}^2 } \; . \nonumber
\eea
Then, integrating $\mu_0$ out of the likelihood ${\cal L}= e^{- \frac{\chi^2_{\rm SN}}{2}}$ we can retrieve
\begin{equation}
\tilde{\chi}^2_{\rm SN}=A- \frac{B^2}{C} + \ln \frac{C}{2 \pi} \, ,
\label{SN8}
\end{equation}
where $\tilde{\chi}^2_{\rm SN}$ has now no dependence on the $\mu_0$ parameter. We have to point out that also in this case we are using a diagonal covariance matrix, because it is not possible to forecast out-of-diagonal terms; this can lead to underestimated errors on cosmological parameters. With BAO we have two correlated measurements to contribute to the total $\chi^2$; these are $H(z) \, r_s$ and $D_A(z)/r_s$. With the Hubble parameter from our phenomenological fit and the angular diameter defined in the previous section, the comoving sound horizon $r_s$ reads
\be
r_s(z_*) =  \frac{1}{H_0} \int_{z_*}^\infty dz'\frac{c_s}{E(z')} = \frac{1}{H_0} \int_0^{a_*} \frac{da'}{a'^2} \frac{c_s}{E(a')}
\ee
where the sound speed is $c_{s} = c /\sqrt{3(1+ \overline{R_b} a)}$, with $\overline{R_b}=31500 \Omega_b h^2 (T_{CMB}/2.7K)^{-4}$ and $T_{CMB}=2.725$ \cite{Fixsen:2009ug}. The comoving sound horizon $r_s(z_*)$ is evaluated at photon-decoupling epoch redshift given by the fitting formula \cite{Hu:1995en}
\bea
z_* &=& 1048 \left[1+0.00124 (\Omega_b h^2)^{-0.738} \right] \\
&\times& \left[1+b_1 (\Omega_m h^2)^{b_2} \right] \; , \nonumber
\eea
with
\bea
b_1 &=& 0.0783 (\Omega_b h^2)^{-0.238} \left[ 1+ 39.5 (\Omega_b h^2)^{0.763} \right]^{-1} \; \; \\
b_2 &=& 0.560 \left[1+ 21.1 (\Omega_b h^2)^{1.81} \right]^{-1} \; ,
\eea
where $\Omega_b$ and $\Omega_b$ are the baryon and matter content of the universe and $h=H_0/100$. The BAO contribution is calculated independently for each redshift, $\chi^2_{BAO}= \sum_i \chi^2_{BAO_i}$, but taking into account the correlation of the magnitudes, each term at each redshift has the form
\be
\chi^2_{BAO_i} = \frac{1}{1-r^2} \left( \frac{\tilde{H}^2_{i}}{\sigma_{\tilde{H}_i}^2} + \frac{\tilde{D}^2_{i}}{\sigma_{\tilde{D}_i}^2} - 2 r \frac{\tilde{H}_{i}}{\sigma_{\tilde{H}_i}} \frac{\tilde{D}_{i}}{\sigma_{\tilde{D}_i}} \right) \, , \\
\ee
where $\tilde{H}_i$ and $\tilde{D}_i$ are the differences between the model predicted and the mock generated measurements:
\bea
\tilde{H}_{i} &=& H(z_i) \, r_s(z_*)  - (H\,r_s)^{mock}_i
, \; \; \\
\tilde{D}_{i} &=& \frac{D_A(z_i)}{r_s(z_*)} - \left( \frac{D_A}{r_s} \right) ^{mock}_i  \, .
\eea
The correlation between the two magnitudes $H\,r_s$ and $D_A/r_s$ in each redshift is fixed as $r=0.4$ \cite{Seo:2007ns}. Since CMB data are not used, SNe data are marginalized over the parameter $H_0$, and BAO data do not give information about it (because $D_{A}/r_s(z_*)$ and $H\, r_s(z_*)$ do not basically depend on it), the parameters $H_0$ and the combination $\Omega_b h^2$ cannot be well constrained. Thus, we also include a Gaussian prior for $H_0$ and $\Omega_b h^2$, with $H_0^{Planck}= 67.51 \pm 0.64$ and for $\Omega_b h^2_{Planck} = 0.02226 \pm 0.00016$ both derived from \textit{Planck} \cite{Ade:2015xua}.

The minimization of the $\chi^2$ function was performed using the MCMC method \cite{Christensen:2001gj,Lewis:2002ah,Trotta:2004qj}, with a Wolfram Mathematica self-developed code based on the Metropolis-Hastings algorithm. In order to see the contribution of each mock data set to the total $\chi^2$, we have also run chains for each data set separately. In this way, we compare the cosmological redshift drift data with those from the other future surveys, and find out whether it will be useful and up to what extent. Moreover, for a round analysis regarding the viability of the Sandage-Loeb test and the performance of the future (mock) surveys, several dark energy scenarios are put to the test.

\subsection{$\Lambda$CDM}

The first model we test is the extremely well-known $\Lambda$CDM model \cite{Carroll:1991mt,Sahni:1999gb}, which has no  degree of freedom in the dark energy equation of state and whose dimensionless Hubble parameter is given by
\be
E_{\Lambda CDM}^2(a) = \Omega_m a^{-3}+\Omega_{r}a^{-4 }+ \Omega_{\Lambda} \, ,
\ee
taking $\Omega_{\Lambda}=1 - \Omega_{m} - \Omega_{r}$ with \cite{Wang2013}
\be
\Omega_{r}= \Omega_{m} \left[ 1 + 2.5 \times 10^4 h^2 \Omega_m (T_{CMB}/2.7)^{-4} \right]^{-1}
\ee
and using $T_{CMB}=2.7255 \, K$ \cite{Fixsen:2009ug}. We enforce $0<\Omega_m<1$, and $0<\Omega_b< \Omega_m<1$ as physical priors, and we do the same for all the rest of models analysed in this paper. The results of the Bayesian analysis for the $\Lambda$CDM model can be seen on Table \ref{tabla_LCDM} and Fig.~\ref{fig:lcdm}.

\subsection{Quiessence}

The second model tested is quiessence \cite{Knop:2003iy,Riess:2004nr}, with a single degree of freedom in the dark energy equation of state parameter (i.e. no redshift dependence). Its dimensionless Hubble parameter is given by
\be
E_{Q}^2(a) = \Omega_m a^{-3}+\Omega_{r}a^{-4 } + \Omega_{\Lambda}a^{-3(1+w)},
\ee
where all the parameters except $w$ are built like in the $\Lambda$CDM model and have the same priors. The  parameter $w$ has the prior $-5<w<0$. This range was chosen decided after having verified that expanding it further has no influence on results. Table \ref{tabla_Q} and Fig.~\ref{fig:q} show the results for quiessence model.

\subsection{Slow-Roll Dark Energy}

We consider another one-parameter dark energy model, coming from the slow-roll dark energy scenario described in \cite{Slepian:2013ug}. Its dimensionless Hubble parameter, taking into account a radiation component \cite{Ade:2015rim,Aubourg:2014yra} is given by
\bea
E_{SR}^2(a) = \Omega_m a^{-3}+\Omega_{r}a^{-4 } + 
+ \Omega_{\Lambda} \left( \frac{a^{-3}}{\Omega_m  a^{-3} +\Omega_r a^{-4} + \Omega_{\Lambda}}\right)^{(\delta w/\Omega_{\Lambda})}\; . \nonumber
\eea
For $\delta w$ we impose a prior of the same width as that of  the parameter $w$ of quiessence, but as $\delta w$ is supposed to have its mean value at $\delta w=0$, we design its prior accordingly. Thus, we take $-2.5 < \delta w < 2.5$. The results for the the slow-roll dark energy model can be found on Table \ref{tabla_SR} and Fig.~\ref{fig:sr}.

\subsection{CPL}

We are also interested in testing models of dark energy whose equation of state parameter $w$ has more than one degree of freedom. As our first two-parameter dark energy model, we take  the CPL model \cite{Chevallier:2000qy,Linder:2002et}, its dimensionless Hubble parameter being
\bea
E_{CPL}^2(a) = \Omega_m a^{-3}+\Omega_{r}a^{-4 } 
+ \Omega_{\Lambda}a^{-3(1+w_0+w_a)} e^{-3 w_a (1-a)}\;, \nonumber
\eea
where all the terms except $w_0$ and $w_a$ are built like in previous models and with the same priors. The parameter $w_0$ has the same prior as $w$ does in quiessence; and we take $-5 < w_a < 5$ for the second parameter. We demand in this case $w_a + w_0 < 0$ in order to have an equation of state for the DE component which is negative in the asymptotic past. Table \ref{tabla_CPL} and Fig.~\ref{fig:cpl} give the results of our Bayesian analysis for the CPL model.

\subsection{Lazkoz-Sendra pivotal Dark Energy}

We consider another model with two parameters for the equation of state for DE \cite{Sendra:2011pt}, which can be understood easily as a perturbative departure from $\Lambda$CDM up to second order in redshift. Even though it is a different parametrization as compared to CPL, it can be also expressed in terms of the parameters $w_0$ and $w_a$ with the same interpretation: $w_0$ is the value of equation of state of the dark energy at present, whereas $w_0+w_a$ is its value in the asymptotic past. Specifically, the Lazkoz-Sendra pivotal dark energy parametrization has the following dimensionless Hubble parameter:
\bea
E_{\rm LS}^2(a) &=& \Omega_m a^{-3} +\Omega_r a^{-4} +\Omega_{\Lambda} X(a) \; , \\
X(a) &=& a^{-3(1+w_0+w_a)} e^{\frac{3}{4}(1-a) \left[ 1+w_0-5 w_a + a(w_a-w_0-1)  \right]} \nonumber
\eea
where all the relative densities $\Omega_i$ are built like in the CPL case, having all the parameters also the same priors as in CPL, including $w_0$ and $w_a$. In the case of the Lazkoz-Sendra model, the results of the Bayesian analysis are shown in Table \ref{tabla_LS} and Fig.~\ref{fig:LS}.

\section{Results and Conclusions}
\label{sec:Results}

In the summary tables for each model, we present the minimum value of $\chi^2$, the constraints for all the free parameters and the reduced  $\chi^2_{red}$. As explained in previous section, the $\chi^2$-minimization is done using different combinations of data sets. In the tables we first show the results from using BAO and SNe separately and those from joining both; then, we move on to present the results from SL only, and then, finally, those for the total SNe+BAO+SL combination. When using SL data, each data set with different observation years is treated separately. In this way, the performance of the cosmological redshift drift data sets can be clearly analysed. For each model we also show the confidence contours for the most interesting cosmological parameters. Each MCMC round is tested for statistical convergence using the method described in \cite{Dunkley2005}.

In the $\Lambda$CDM scenario we find that the cosmological redshift drift data provide remarkably good constraints on  $\Omega_m$: when those data are used alone we get standard deviations on $\Omega_m$ which are $2-3$ times smaller that those from the SNe+BAO combination. Considering the broad priors taken for $\Omega_m$ in all cases, and the negligible correlation between the Hubble constant $h$ and $\Omega_m$,\footnote{As the major axis of the contours are typically aligned with the axes of each parameters in the parameter space.} we conclude that the result for the matter density $\Omega_m$ is not influenced by any prior and is solely given by the data.

Indeed, the SL data sets always do better in constraining $\Omega_m$ than the SNe data, and depending on the model and on the years of observation, even better than the BAO data set. Once we combine the SL data set with the other two, the cosmological redshift drift is still helpful, even though the BAO+SNe data set already improves greatly the constrains in the parameter space. In general, it is clear that the cosmological redshift drift data helps considerably to constrain the parameter $\Omega_m$ in all the models.

Regarding the dark energy parameters, we can observe that for most of the cases, the $24$ years of observation for SL is not enough to constrain them properly, as it can be clearly seen for example from the contours of the parameters $w_0$ and $w_a$ in Figs.~\ref{fig:cpl}~-~\ref{fig:LS}. With $28$ years SL data, the $1 \sigma$ regions improve noticeable, and with $32$ years of observation both $1\sigma$ and $2\sigma$ regions are well constrained for all the DE parameters. The best example is in Fig.~\ref{fig:LS}, as stated, but similar behaviour can be appreciated in the rest of the models. Besides, it is clear that increasing the observation years improves the overall constraining ability of the cosmological redshift drift data sets. It is worth to note that in all these cases, the contours of the SL data set are almost perpendicular to the contours of the SNe and BAO data sets, thus showing a great complementarity between SL and the rest of the data sets \cite{Coe:2009xf}, as for example in the $\Omega_m-w$ plane for the quiessence model, Fig.~\ref{fig:q}, or for the dynamical dark energy models, Figs.~\ref{fig:sr}~-~\ref{fig:cpl}~-~\ref{fig:LS}. This is very important, because it means that even the cosmological redshift drift data set with the lowest observation period, noticeably contributes to improved dark energy insights when used as cosmological probe together with other kind of observations.

\begin{figure}[h!]
\begin{minipage}{\columnwidth}
	\includegraphics[width=0.5\textwidth]{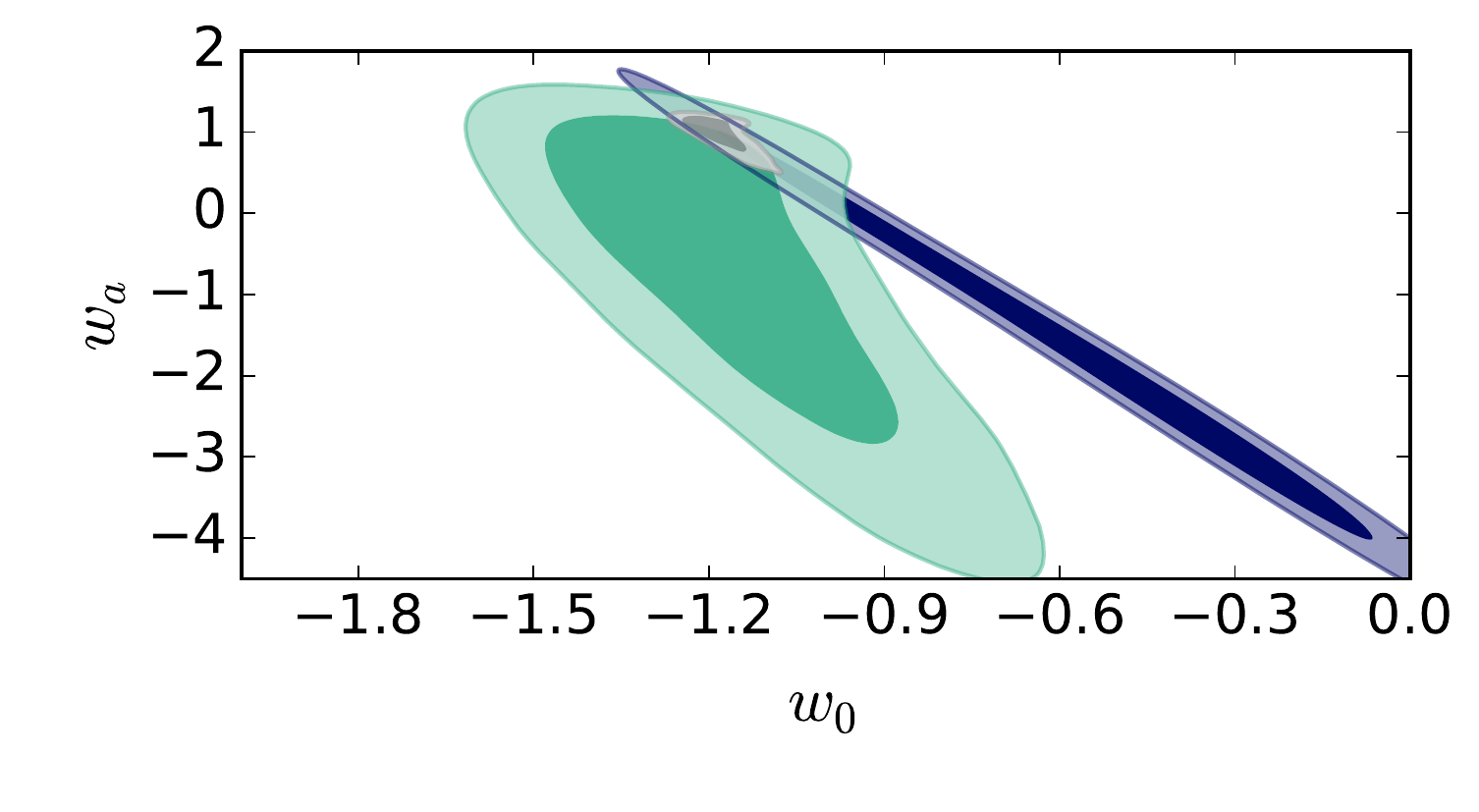}\includegraphics[width=0.5\textwidth]{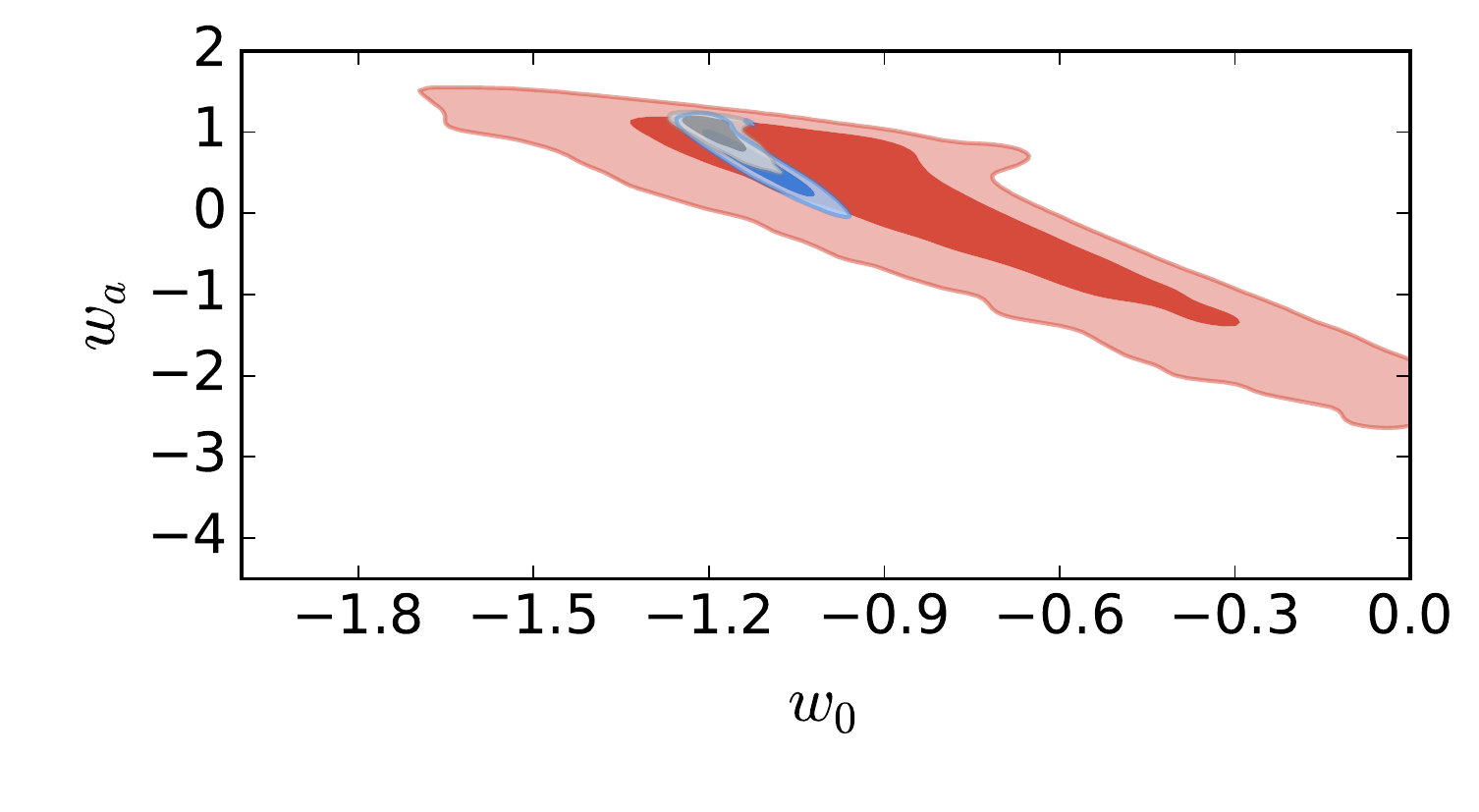}
	\caption{Contours in the $w_0 - w_a$ plane for CPL; solid contours are for $1\sigma$ regions and clear contours are for $2\sigma$ regions. Left panel: purple is for the BAO data; green for SNe and grey for SNe+BAO. Right panel: red is for $32$ years SL data; grey for SNe+BAO; blue for SNe+BAO+SL.}
	\label{fig:w0-wa}       
\end{minipage}
\end{figure}

However, if one focuses on the $w_0$ and $w_a$ parameters, two things can be noted: first, that the best fit for the SNe+BAO case is completely different from the values derived from SNe and BAO only analysis (this is more evident for $w_{a}$ than $w_0$); second, the errors on the $w_0$ and $w_a$ parameters slightly increase when adding cosmological redshift drift data to the SNe+BAO data. Both trends might have an explanation. For what concerns the first one, if we look at the left panel of Fig.~\ref{fig:w0-wa} (this is for the CPL case, but it holds true for the LS model as well), we can see how unsatisfactorily the SNe and BAO contours overlap: the borders of the $1\sigma$ confidence levels show a small overlap in a region which is far from the best fit expected for each of them when considered separately. This reduces the constraints on the parameters in a considerable way and shifts the best fit estimations (not only in $w_{a}$, but also on $\Omega_m$). But note also that this behavior is somehow expected and might be counter-productive in the future, as explained in \cite{tension}. Anyway, we must also remember we are working with mock data, not real ones, and the potential future goodness of the joint use of SNe and BAO at the present, and maybe in the near future, is not put at stake. Moreover, we have to remember that in order to gain more insights into a dynamical dark energy model, we need to improve the number and the quality of data at high redshift; that is the reason behind pushing SNe observations to higher redshifts \cite{Salzano2013}, for example, or employing BAO data at $z\sim 2$. But the strongest hints about the dynamical nature of the dark energy might come from data like SL, which are able to cover a larger and deeper redshift range. And the second issue discussed above should be exactly connected to this: if we check again the right panel of Fig.~\ref{fig:w0-wa}, we can see how the SL data set alone, which should be more sensitive to a dynamical dark energy, determines a consistent shift in the parameter $w_{0}$ with respect to the SNe+BAO case but with smaller uncertainty with respect to both SNe and BAO data separately, which eventually ends in a slightly large error for this parameter for the total SNe+BAO+SL sample.

In the case of models with a single DE parameter, that is, whose equation of state is fixed during time, high redshift SL data are also helpful. In the extreme case when SL data are added to the SNe+BAO data set, even the SL data with lowest observational period help constrain the single parameter of DE. However, it is also remarkable how every data set, separately, constrains the single DE parameter to a different value. Taking into account that the redshift range of each data set is quite different, the fact that separately they measure a different value for the parameter could be an evidence for a time evolution in the equation of state of DE. This is a clear example of another application for the SL observation, where its high redshift data could easily test the time evolution of the equation of state of DE once compared to the results of other data sets coming from different sources.

A lot of what has been stated above can be easily inferred upon closer examination of the various contours plots. However, these plots are more useful for analyzing the correlation between different parameters. As stated previously, in most of the contour plots, a different correlation angle can be seen for the cosmological redshift drift data comparing the other data sets. Thus, it clearly emerges that SL data sets will be of utmost importance in breaking degeneracies among cosmological parameters. Besides, considering the high redshift data that will be available thanks to cosmological redshift drift, we conclude that it can be a cosmic observable much worth to consider. \\

\begin{acknowledgements}
	R.L. and I.L. were supported by the Spanish Ministry of Economy and Competitiveness through research projects No. FIS2014-57956-P (comprising FEDER funds) and also by the Basque Government through research project No. GIC17/116-IT956-16.
	I.L. acknowledges financial support from the University of the Basque Country UPV/EHU PhD grant No. 750/2014.
	V.S. is funded by the Polish National Science Center Grant No. DEC-2012/06/A/ST2/00395.
	This article is based upon work from COST Action CA15117 (CANTATA), supported by COST (European Cooperation in Science and Technology).
\end{acknowledgements}

\appendix


\section{Results for the $\Lambda$CDM model}

\begin{table}[h]
	\caption{Parameter results of the $\Lambda$CDM model.}
	\label{tabla_LCDM}       
	\begin{tabular}{lllllll}
		\hline\noalign{\smallskip}
		\bf{Data Set} & & $\bf{h}$ & $\bf{\Omega_m}$ & $\bf{\Omega_b}$ & $\bf{\chi^2_{min}}$ & $\bf{\chi^2_{red}}$\\
		\noalign{\smallskip}\hline\noalign{\smallskip}
		BAO & & $0.689_{-0.002}^{+0.002}$ & $0.335^{+0.008}_{-0.008}$ & $0.0467^{+0.0004}_{-0.0004}$ & $9.47$ & $0.861$\\
		SNe & & $0.675_{-0.007}^{+0.006}$ & $0.301^{+0.008}_{-0.008}$ & & $1387.41$ & $0.510$\\
		\hline
		SNe+BAO & & $0.689_{-0.002}^{+0.002}$ & $0.324_{-0.006}^{+0.006}$ & $0.0472_{-0.0004}^{+0.0004}$ & $1402.07$ & $0.513$\\
		\hline
		SL(24y) & & $0.674_{-0.006}^{+0.006}$ & $0.328^{+0.003}_{-0.003}$ & & $14.51$ & $0.538$\\
		SNe+BAO+SL(24y) & & $0.689_{-0.002}^{+0.002}$ & $0.324^{+0.003}_{-0.002}$ & $0.0467^{+0.0004}_{-0.0004}$ & $1417.82$ & $0.513$\\
		\hline
		SL(28y) & & $0.673_{-0.006}^{+0.006}$ & $0.328^{+0.003}_{-0.003}$ & & $19.73$ & $0.731$\\
		SNe+BAO+SL(28y) & & $0.689_{-0.002}^{+0.002}$ & $0.325^{+0.002}_{-0.002}$ & $0.0467^{+0.0004}_{-0.0004}$ & $1423.49$ & $0.515$\\
		\hline
		SL(32y) & & $0.673_{-0.006}^{+0.006}$ & $0.328^{+0.002}_{-0.002}$ & & $25.75$ & $0.954$\\
		SNe+BAO+SL(32y) & & $0.689_{-0.002}^{+0.002}$ & $0.325^{+0.002}_{-0.002}$ & $0.0468^{+0.0004}_{-0.0004}$ & $1430.03$ & $0.518$\\
		\noalign{\smallskip}\hline
	\end{tabular}
\end{table}

\begin{figure}[h]
\centering	
	\includegraphics[width=\textwidth]{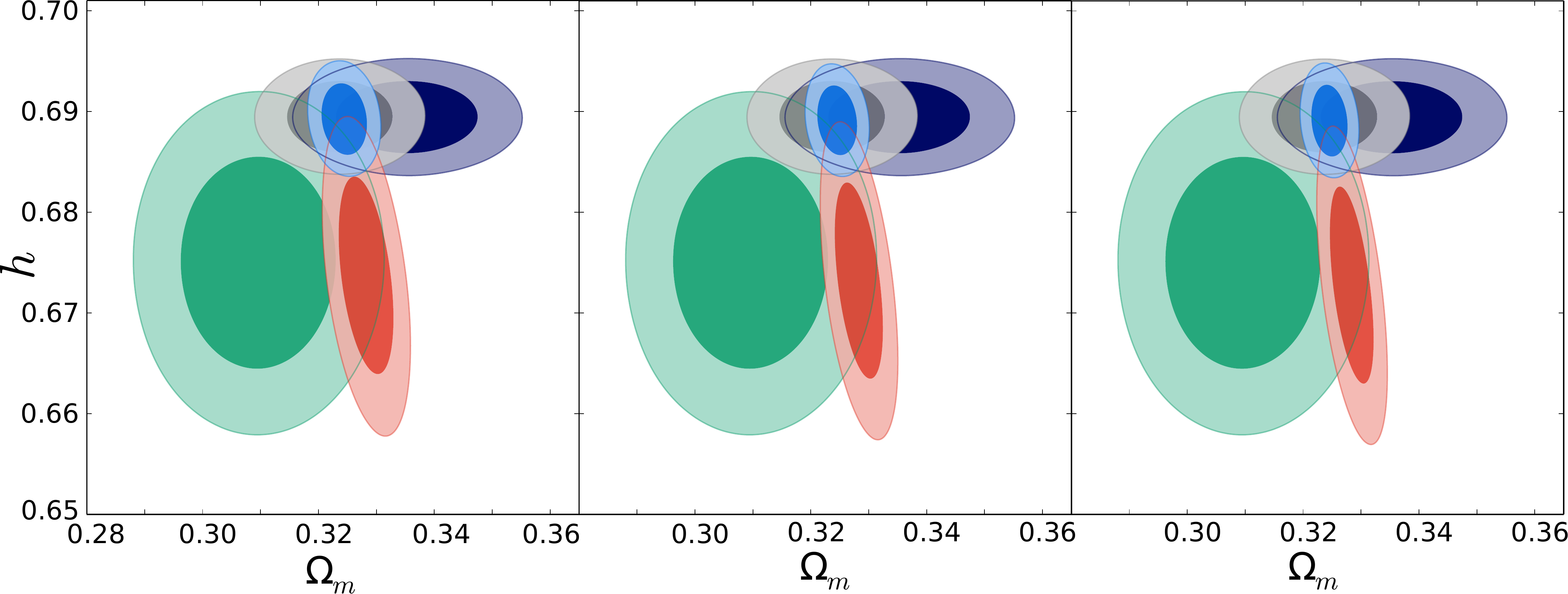}
	\caption{$\Lambda$CDM model; solid contours limit $1\sigma$ regions and clear contours $2\sigma$ region.  Purple for the BAO, green for SNe and red for SL data set, grey for SNe+BAO and blue SL+BAO+SNe. First set (left) for 24 years, second (middle) for 28 and third (right) for 32 years.}
	\label{fig:lcdm}       
\end{figure}

\newpage
\section{Results for the Quiessence model}

\begin{table}[h]
	\caption{Parameter results of the Quiessence model.}
	\label{tabla_Q}       
	\begin{tabular}{llllllll}
		\hline\noalign{\smallskip}
		\bf{Data Set} & & $\bf{h}$ & $\bf{\Omega_m}$ & $\bf{\Omega_b}$ & $\bf{w}$ & $\bf{\chi^2_{min}}$ & $\bf{\chi^2_{red}}$\\
		\noalign{\smallskip}\hline\noalign{\smallskip}
		BAO & & $0.677_{-0.006}^{+0.006}$ & $0.336^{+0.008}_{-0.008}$ & $0.0485^{+0.0010}_{-0.0009}$ & $-0.948^{+0.024}_{-0.025}$ & $5.10$ & $0.510$\\
		SNe & & $0.675_{-0.007}^{+0.006}$ & $0.341^{+0.013}_{-0.015}$ & & $-1.244^{+0.123}_{-0.122}$ & $1383.13$ & $0.509$\\
		\hline
		SNe+BAO & & $0.686_{-0.006}^{+0.006}$ & $0.323_{-0.006}^{+0.006}$ & $0.0472_{-0.0008}^{+0.0008}$ & $-0.987_{-0.023}^{+0.022}$ & $1401.74$ & $0.513$\\
		\hline
		SL(24y) & & $0.674_{-0.007}^{+0.006}$ & $0.323^{+0.011}_{-0.011}$ & &  $-0.888^{+0.141}_{-0.660}$ & $12.73$ & $0.490$\\
		SNe+BAO+SL(24y) & & $0.685_{-0.005}^{+0.005}$ & $0.324^{+0.003}_{-0.002}$ & $0.0474^{+0.0008}_{-0.0008}$ & $-0.982^{+0.020}_{-0.021}$ & $1417.07$ & $0.513$\\
		\hline
		SL(28y) & & $0.674_{-0.007}^{+0.006}$ & $0.321^{+0.009}_{-0.009}$ & & $-0.845^{+0.102}_{-0.176}$ & $17.32$ & $0.666$\\
		SNe+BAO+SL(28y) & & $0.684_{-0.005}^{+0.005}$ & $0.325^{+0.002}_{-0.002}$ & $0.0475^{+0.0008}_{-0.0008}$ & $-0.979^{+0.020}_{-0.021}$ & $1422.49$ & $0.515$\\
		\hline
		SL(32y) & & $0.674_{-0.006}^{+0.007}$ & $0.320^{+0.007}_{-0.008}$ & & $-0.830^{+0.084}_{-0.119}$ & $22.62$ & $0.870$\\
		SNe+BAO+SL(32y) & & $0.684_{-0.005}^{+0.005}$ & $0.325^{+0.002}_{-0.002}$ & $0.0476^{+0.0008}_{-0.0008}$ & $-0.977^{+0.020}_{-0.020}$ & $1428.75$ & $0.517$\\
		\noalign{\smallskip}\hline
	\end{tabular}
\end{table}

\begin{figure*}[h]
\centering
	\includegraphics[width=\textwidth]{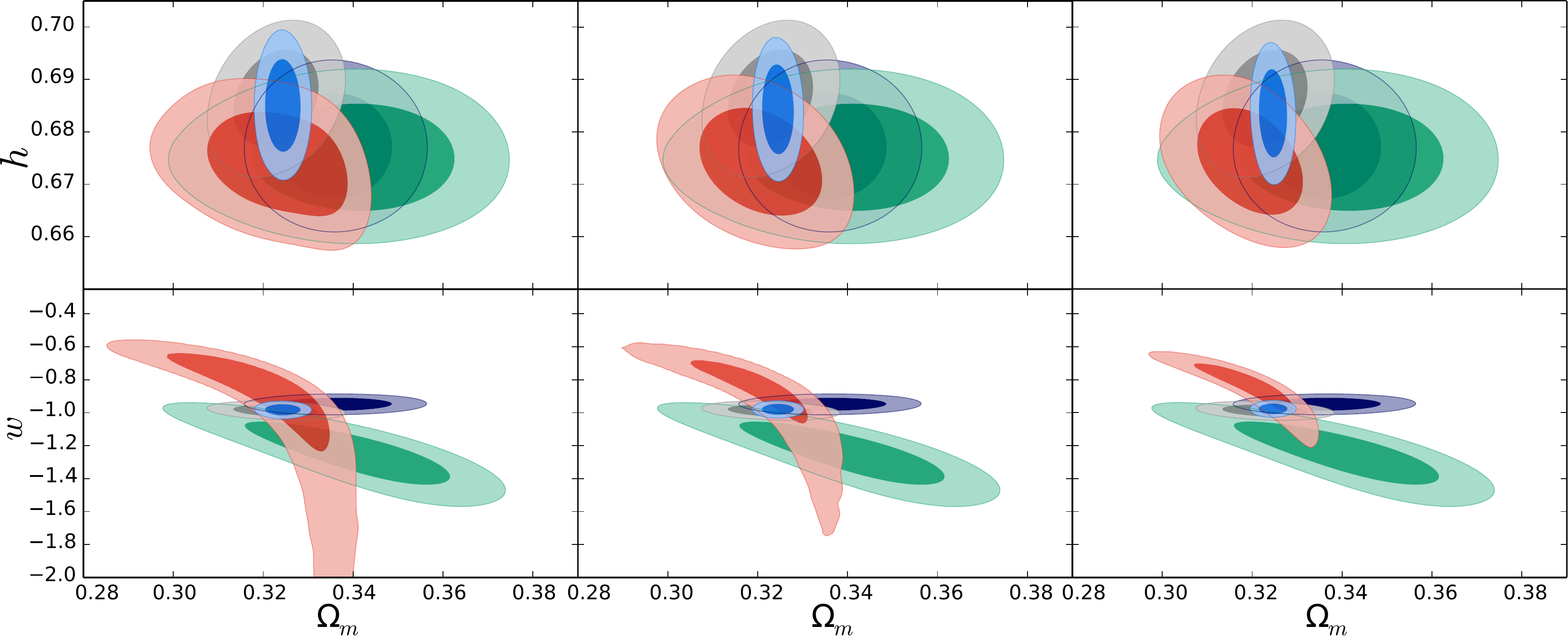}
	\caption{Quiessence model; solid contours limit $1\sigma$ regions and clear contours $2\sigma$ region.  Purple for the BAO, green for SNe and red for SL data set, grey for SNe+BAO and blue SL+BAO+SNe. First set (left) for 24 years, second (middle) for 28 and third (right) for 32 years.}
	\label{fig:q}       
\end{figure*}

\newpage
\section{Results for the Slow-Roll model}

\begin{table}[h]
	\caption{Parameter results of the Slow-Roll model.}
	\label{tabla_SR}       
	\begin{tabular}{llllllll}
		\hline\noalign{\smallskip}
		\bf{Data Set} & & $\bf{h}$ & $\bf{\Omega_m}$ & $\bf{\Omega_b}$ & $\bf{\delta w}$ & $\bf{\chi^2_{min}}$ & $\bf{\chi^2_{red}}$\\
		\noalign{\smallskip}\hline\noalign{\smallskip}
		BAO & & $0.678_{-0.006}^{+0.006}$ & $0.341_{-0.008}^{+0.009}$ & $0.0485_{-0.0009}^{+0.0010}$ & $0.074_{-0.035}^{+0.036}$ & $5.20$ & $0.520$\\
		SNe & & $0.675_{-0.006}^{+0.006}$ & $0.330_{-0.012}^{+0.010}$ & & $-0.260_{-0.138}^{+0.128}$ & $1383.10$ & $0.509$\\
		\hline
		SNe+BAO & & $0.688_{-0.006}^{+0.005}$ & $0.324_{-0.006}^{+0.006}$ & $0.0468_{-0.0008}^{+0.0008}$ & $0.005_{-0.030}^{+0.030}$ & $1402.05$ & $0.513$\\
		\hline
		SL(24y) & & $0.674_{-0.006}^{+0.006}$ & $0.324_{-0.007}^{+0.007}$ & &  $0.231_{-0.539}^{+0.252}$ & $13.12$ & $0.504$\\
		SNe+BAO+SL(24y) & & $0.687_{-0.005}^{+0.005}$ & $0.325_{-0.002}^{+0.002}$ & $0.0471_{-0.0008}^{+0.0008}$ & $0.014_{-0.029}^{+0.028}$ & $1417.62$ & $0.513$\\
		\hline
		SL(28y) & & $0.674_{-0.006}^{+0.006}$ & $0.323_{-0.006}^{+0.006}$ & & $0.282_{-0.330}^{+0.199}$ & $17.85$ & $0.686$\\
		SNe+BAO+SL(28y) & & $0.686_{-0.005}^{+0.005}$ & $0.325_{-0.002}^{+0.002}$ & $0.0472_{-0.0008}^{+0.0008}$ & $0.017_{-0.027}^{+0.028}$ & $1423.21$ & $0.515$\\
		\hline
		SL(32y) & & $0.674_{-0.006}^{+0.006}$ & $0.323_{-0.005}^{+0.005}$ & & $0.306_{-0.254}^{+0.171}$ & $23.31$ & $0.897$\\
		SNe+BAO+SL(32y) & & $0.686_{-0.005}^{+0.005}$ & $0.325_{-0.002}^{+0.002}$ & $0.0473_{-0.0007}^{+0.0008}$ & $0.021_{-0.029}^{+0.029}$ & $1429.59$ & $0.518$\\
		\noalign{\smallskip}\hline
	\end{tabular}
\end{table}

\begin{figure*}[h]
\centering
	\includegraphics[width=\textwidth]{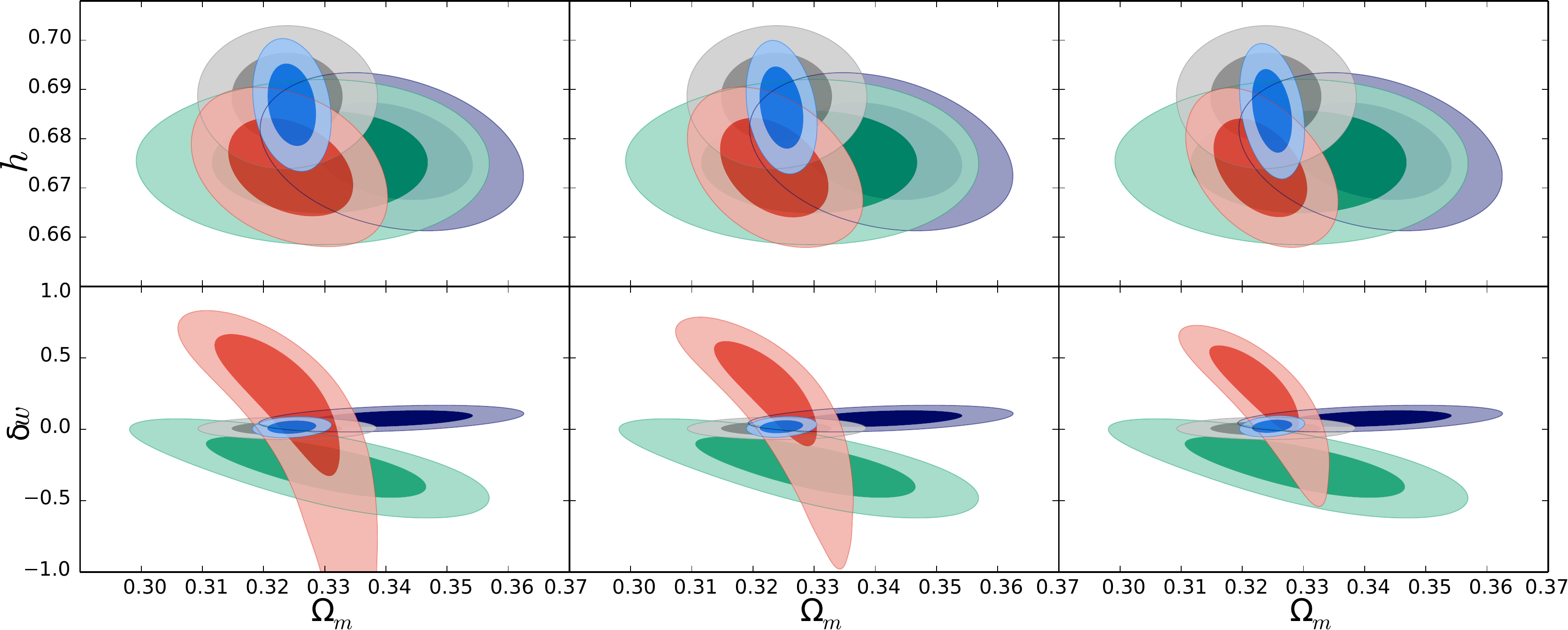}
	\caption{Slow-Roll model; solid contours limit $1\sigma$ regions and clear contours $2\sigma$ region.  Purple for the BAO, green for SNe and red for SL data set, grey for SNe+BAO and blue SL+BAO+SNe. First set (left) for 24 years, second (middle) for 28 and third (right) for 32 years.}
	\label{fig:sr}       
\end{figure*}

\newpage
\section{Results for the CPL model}

\begin{table}[h]
	\caption{Parameter results of the CPL model.}
	\label{tabla_CPL}       
	\begin{tabular}{lllllllll}
		\hline\noalign{\smallskip}
		\bf{Data Set} & & $\bf{h}$ & $\bf{\Omega_m}$ & $\bf{\Omega_b}$ & $\bf{w_0}$ & $\bf{w_a}$ & $\bf{\chi^2_{min}}$ & $\bf{\chi^2_{red}}$\\
		\noalign{\smallskip}\hline\noalign{\smallskip}
		BAO & & $0.677_{-0.007}^{+0.006}$ & $0.395^{+0.035}_{-0.075}$ & $0.0486^{+0.0010}_{-0.0010}$ & $-0.558^{+0.311}_{-0.467}$ & $-1.722^{+2.035}_{-1.457}$ & $3.37$ & $0.374$\\
		SNe & & $0.675_{-0.006}^{+0.006}$ & $0.356^{+0.030}_{-0.038}$ & & $-1.165^{+0.203}_{-0.182}$ & $-0.555^{+1.232}_{-1.708}$ & $1383.07$ & $0.509$\\
		\hline
		SNe+BAO & & $0.678_{-0.006}^{+0.006}$ & $0.281^{+0.011}_{-0.011}$ & $0.0484^{+0.0010}_{-0.0009}$ & $-1.187^{+0.040}_{-0.032}$ & $1.022^{+0.119}_{-0.195}$ & $1386.45$ & $0.508$\\
		\hline
		SL(24y) & & $0.674_{-0.006}^{+0.007}$ & $0.328^{+0.008}_{-0.017}$ & & $-1.026^{+0.498}_{-1.518}$ & $-0.001^{+1.044}_{-1.584}$ & $12.19$ & $0.488$\\
		SNe+BAO+SL(24y) & & $0.684_{-0.006}^{+0.006}$ & $0.314_{-0.008}^{+0.006}$ & $0.0476_{-0.0008}^{+0.0009}$ & $-1.117_{-0.066}^{+0.066}$ & $0.602_{-0.286}^{+0.286}$ & $1412.27$ & $0.512$\\
		\hline
		SL(28y) & & $.674_{-0.006}^{+0.006}$ & $0.324^{+0.010}_{-0.017}$ & & $-0.937^{+0.423}_{-0.627}$ & $-0.021^{+0.869}_{-1.541}$ & $16.56$ & $0.662$\\
		SNe+BAO+SL(28y) & & $0.683_{-0.006}^{+0.006}$ & $0.315_{-0.008}^{+0.006}$ & $0.0476_{-0.0008}^{+0.0009}$ & $-1.113_{-0.068}^{+0.066}$ & $0.590_{-0.276}^{+0.293}$ & $1417.52$ & $0.513$\\
		\hline
		SL(32y) & & $0.674_{-0.006}^{+0.006}$ & $0.319^{+0.010}_{-0.019}$ & & $-0.895^{+0.372}_{-0.256}$ & $0.240^{+0.637}_{-1.147}$ & $21.65$ & $0.866$\\
		SNe+BAO+SL(32y) & & $0.683_{-0.006}^{+0.006}$ & $0.315_{-0.008}^{+0.006}$ & $0.0477_{-0.0008}^{+0.0009}$ & $-1.120_{-0.062}^{+0.066}$ & $0.628_{-0.278}^{+0.297}$ & $1423.36$ & $0.516$\\
		\noalign{\smallskip}\hline
	\end{tabular}
\end{table}

\begin{figure*}[h]
\centering
	\includegraphics[width=\textwidth]{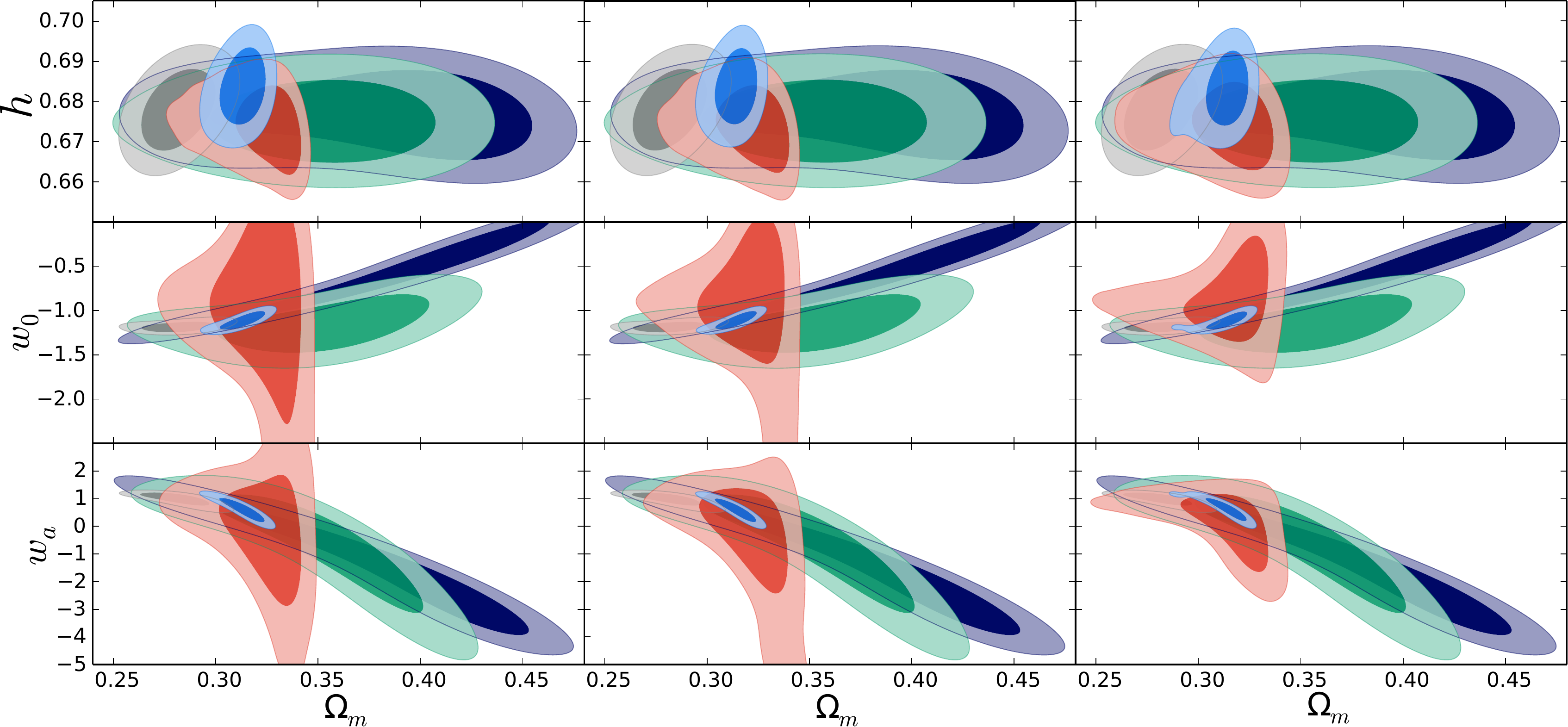}
	\caption{CPL model; solid contours limit $1\sigma$ regions and clear contours $2\sigma$ region.  Purple for the BAO, green for SNe and red for SL data set, grey for SNe+BAO and blue SL+BAO+SNe. First set (left) for 24 years, second (middle) for 28 and third (right) for 32 years.}
	\label{fig:cpl}       
\end{figure*}

\newpage
\section{Results for the Lazkoz-Sendra pivotal Dark Energy model}

\begin{table}[h]
	\caption{Parameter results of the Lazkoz-Sendra pivotal model.}
	\label{tabla_LS}       
	\begin{tabular}{lllllllll}
		\hline\noalign{\smallskip}
		\bf{Data Set} & & $\bf{h}$ & $\bf{\Omega_m}$ & $\bf{\Omega_b}$ & $\bf{w_0}$ & $\bf{w_a}$ & $\bf{\chi^2_{min}}$ & $\bf{\chi^2_{red}}$\\
		\noalign{\smallskip}\hline\noalign{\smallskip}
		BAO & & $0.677_{-0.006}^{+0.006}$ & $0.391_{-0.065}^{+0.034}$ & $0.0485_{-0.0009}^{+0.0009}$ & $-0.661_{-0.333}^{+0.244}$ & $-2.114_{-1.820}^{+2.401}$ & $3.33$ & $0.370$\\
		SNe & & $0.675_{-0.007}^{+0.006}$ & $0.361_{-0.031}^{+0.023}$ & & $-1.170_{-0.143}^{+0.150}$ & $-1.063_{-2.004}^{+1.565}$ & $1383.13$ & $0.509$\\
		\hline
		SNe+BAO & & $0.680_{-0.006}^{+0.006}$ & $0.295_{-0.008}^{+0.009}$ & $0.0481_{-0.0009}^{+0.0009}$ & $-1.093_{-0.025}^{+0.028}$ & $0.934_{-0.196}^{+0.112}$ & $1388.46$ & $0.508$\\
		\hline
		SL(24y) & & $0.673_{-0.006}^{+0.007}$ & $0.331_{-0.014}^{+0.005}$ & & $-1.056_{-2.390}^{+0.510}$ & $-0.616_{-2.186}^{+1.416}$ & $12.53$ & $0.501$\\
		SNe+BAO+SL(24y) & & $0.683_{-0.006}^{+0.006}$ & $0.311_{-0.006}^{+0.006}$ & $0.0477_{-0.0008}^{+0.0008}$ & $-1.088_{-0.029}^{+0.037}$ & $0.826_{-0.278}^{+0.185}$ & $1408.54$ & $0.510$\\
		\hline
		SL(28y) & & $0.674_{-0.007}^{+0.007}$ & $0.325_{-0.015}^{+0.009}$ & & $-0.884_{-0.449}^{+0.349}$ & $-0.173_{-1.775}^{+0.917}$ & $16.77$ & $0.671$\\
		SNe+BAO+SL(28y) & & $0.682_{-0.006}^{+0.005}$ & $0.312_{-0.006}^{+0.006}$ & $0.0477_{-0.0008}^{+0.0009}$ & $-1.089_{-0.029}^{+0.037}$ & $0.839_{-0.279}^{+0.177}$ & $1413.75$ & $0.512$\\
		\hline
		SL(32y) & & $0.674_{-0.006}^{+0.006}$ & $0.322_{-0.015}^{+0.009}$ & & $-0.848_{-0.227}^{+0.335}$ & $-0.080_{-1.498}^{+0.815}$ & $21.93$ & $0.877$\\
		SNe+BAO+SL(32y) & & $0.682_{-0.005}^{+0.005}$ & $0.312_{-0.006}^{+0.005}$ & $0.0479_{-0.0008}^{+0.0008}$ & $-1.087_{-0.028}^{+0.038}$ & $0.835_{-0.274}^{+0.177}$ & $1419.53$ & $0.514$\\
		\noalign{\smallskip}\hline
	\end{tabular}
\end{table}

\begin{figure*}[h]
\centering	
	\includegraphics[width=\textwidth]{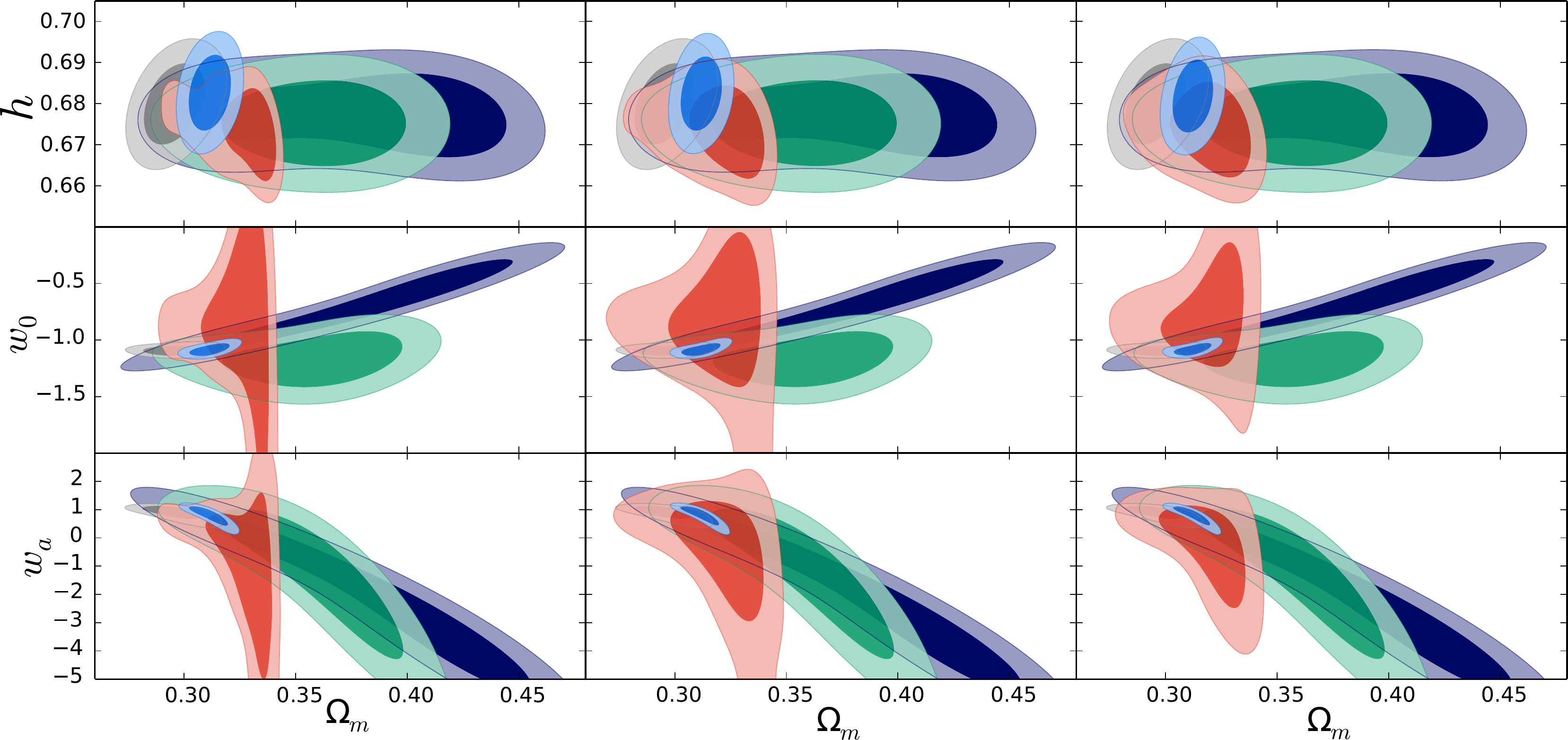}
	\caption{Lazkoz-Sendra pivotal model; solid contours limit $1\sigma$ regions and clear contours $2\sigma$ region.  Purple for the BAO, green for SNe and red for SL data set, grey for SNe+BAO and blue SL+BAO+SNe. First set (left) for 24 years, second (middle) for 28 and third (right) for 32 years.}
	\label{fig:LS}       
\end{figure*}

\bibliography{biblio}

\begin{thebibliography}{10}
\providecommand{\url}[1]{{#1}}
\providecommand{\urlprefix}{URL }
\expandafter\ifx\csname urlstyle\endcsname\relax
  \providecommand{\doi}[1]{DOI \discretionary{}{}{}#1}\else
  \providecommand{\doi}{DOI \discretionary{}{}{}\begingroup
  \urlstyle{rm}\Url}\fi

\bibitem{Carroll:2000fy}
S.M. Carroll, Living Rev. Rel. \textbf{4}, 1 (2001)

\bibitem{Barboza:2015rsa}
E.M. Barboza, R.C. Nunes, E.M.C. Abreu, J.A. Neto, Phys. Rev. \textbf{D92}(8),
  083526 (2015)

\bibitem{Li:2012dt}
M.~Li, X.D. Li, S.~Wang, Y.~Wang, Front. Phys. China \textbf{8}, 828 (2013)

\bibitem{Kunz:2012aw}
M.~Kunz, Comptes Rendus Physique \textbf{13}, 539 (2012)

\bibitem{Copeland:2006wr}
E.J. Copeland, M.~Sami, S.~Tsujikawa, Int. J. Mod. Phys. \textbf{D15}, 1753
  (2006)

\bibitem{Bamba:2012cp}
K.~Bamba, S.~Capozziello, S.~Nojiri, S.D. Odintsov, Astrophys. Space Sci.
  \textbf{342}, 155 (2012)

\bibitem{Mortonson:2013zfa}
M.J. Mortonson, D.H. Weinberg, M.~White, arXiv:1401.0046 [astro-ph.CO]  (2014)

\bibitem{Battye:2015hza}
R.A. Battye, B.~Bolliet, J.A. Pearson, Phys. Rev. \textbf{D93}(4), 044026
  (2016)

\bibitem{Riess:1998cb}
A.G. Riess, et~al., Astron. J. \textbf{116}, 1009 (1998)

\bibitem{Perlmutter:1998np}
S.~Perlmutter, et~al., Astrophys. J. \textbf{517}, 565 (1999)

\bibitem{Suzuki:2011hu}
N.~Suzuki, D.~Rubin, C.~Lidman, G.~Aldering, R.~Amanullah, et~al., Astrophys.
  J. \textbf{746}, 85 (2012)

\bibitem{Beutler:2011hx}
F.~Beutler, C.~Blake, M.~Colless, D.H. Jones, L.~Staveley-Smith, L.~Campbell,
  Q.~Parker, W.~Saunders, F.~Watson, Mon. Not. Roy. Astron. Soc. \textbf{416},
  3017 (2011)

\bibitem{Ade:2015xua}
P.A.R. Ade, et~al., Astron. Astrophys. \textbf{594}, A13 (2016)

\bibitem{cmb2}
Y.~Wang, M.~Dai, Phys. Rev. \textbf{D94}(8), 083521 (2016)

\bibitem{sdss1}
D.J. Eisenstein, et~al., Astrophys. J. \textbf{633}, 560 (2005)

\bibitem{sdss2}
S.~Alam, et~al., Astrophys. J. Suppl. \textbf{219}(1), 12 (2015)

\bibitem{bao2}
S.~Alam, et~al., Mon. Not. Roy. Astron. Soc. \textbf{470}(3), 2617 (2017)

\bibitem{bao3}
A.~Font-Ribera, et~al., JCAP \textbf{1405}, 027 (2014)

\bibitem{Blake2012}
C.~Blake, et~al., Mon. Not. Roy. Astron. Soc. \textbf{425}, 405 (2012)

\bibitem{Weinberg:2012es}
D.H. Weinberg, M.J. Mortonson, D.J. Eisenstein, C.~Hirata, A.G. Riess, E.~Rozo,
  Phys.Rept. \textbf{530}, 87 (2013)

\bibitem{Bartelmann:1999yn}
M.~Bartelmann, P.~Schneider, Phys. Rept. \textbf{340}, 291 (2001)

\bibitem{Sandage:1962}
S.~Allan, Astrophys. J. \textbf{136}, 319 (1962)

\bibitem{Loeb:1998bu}
A.~Loeb, Astrophys. J. \textbf{499}, L111 (1998)

\bibitem{Laureijs:2011gra}
R.~Laureijs, et~al., Euclid Definition Study Report, arXiv:1110.3193
  [astro-ph.CO]  (2011)

\bibitem{Spergel:2013tha}
D.~Spergel, et~al., WFIRST-AFTA Final Report, arXiv:1305.5422 [astro-ph.IM]
  (2013)

\bibitem{SKA2015a}
SKA Level 0 Science Requirements, document SKA-TEL-SKO-0000007  (2015)

\bibitem{Killedar:2009xw}
M.~Killedar, G.F. Lewis, Mon. Not. Roy. Astron. Soc. \textbf{402}, 650 (2010)

\bibitem{Liske:2008ph}
J.~Liske, et~al., Mon. Not. Roy. Astron. Soc. \textbf{386}, 1192 (2008).
\newblock \doi{10.1111/j.1365-2966.2008.13090.x}

\bibitem{Klockner:2015rqa}
H.R. Klöckner, D.~Obreschkow, C.~Martins, A.~Raccanelli, D.~Champion, A.L.
  Roy, A.~Lobanov, J.~Wagner, R.~Keller, PoS \textbf{AASKA14}, 027 (2015)

\bibitem{Zhang:2013fqa}
M.J. Zhang, W.B. Liu, Res. Astron. Astrophys. \textbf{13}, 1397 (2013)

\bibitem{Vielzeuf:2012zd}
P.E. Vielzeuf, C.J.A.P. Martins, Phys. Rev. \textbf{D85}, 087301 (2012)

\bibitem{Corasaniti:2007bg}
P.S. Corasaniti, D.~Huterer, A.~Melchiorri, Phys. Rev. \textbf{D75}, 062001
  (2007).
\newblock \doi{10.1103/PhysRevD.75.062001}

\bibitem{Moraes:2011vq}
B.~Moraes, D.~Polarski, Phys. Rev. \textbf{D84}, 104003 (2011)

\bibitem{Koksbang:2015ctu}
S.M. Koksbang, S.~Hannestad, JCAP \textbf{1601}, 009 (2016)

\bibitem{Yoo:2010hi}
C.M. Yoo, T.~Kai, K.i. Nakao, Phys. Rev. \textbf{D83}, 043527 (2011)

\bibitem{Araujo:2010ir}
M.E. Araujo, W.R. Stoeger, Phys. Rev. \textbf{D82}, 123513 (2010)

\bibitem{Melia:2016bnb}
F.~Melia, Mon. Not. Roy. Astron. Soc. \textbf{463}, L61 (2016)

\bibitem{Denkiewicz:2014kna}
T.~Denkiewicz, M.P. Dabrowski, C.J.A.P. Martins, P.E. Vielzeuf, Phys. Rev.
  \textbf{D89}(8), 083514 (2014)

\bibitem{Zhang:2010im}
J.~Zhang, L.~Zhang, X.~Zhang, Phys. Lett. \textbf{B691}, 11 (2010)

\bibitem{Banerjee:2015ala}
S.~Banerjee, N.~Jayswal, T.P. Singh, Phys. Rev. \textbf{D92}(8), 084026 (2015)

\bibitem{Mishra:2012vi}
P.~Mishra, M.N. Celerier, T.P. Singh, Phys. Rev. \textbf{D86}, 083520 (2012)

\bibitem{Balcerzak:2012bv}
A.~Balcerzak, M.P. Dabrowski, Phys. Rev. \textbf{D87}(6), 063506 (2013)

\bibitem{Balcerzak:2013kha}
A.~Balcerzak, M.P. Dabrowski, Phys. Lett. \textbf{B728}, 15 (2014)

\bibitem{Guo:2015gpa}
R.Y. Guo, X.~Zhang, Eur. Phys. J. \textbf{C76}(3), 163 (2016)

\bibitem{Geng:2014hoa}
J.J. Geng, J.F. Zhang, X.~Zhang, JCAP \textbf{1407}, 006 (2014)

\bibitem{Geng:2014ypa}
J.J. Geng, J.F. Zhang, X.~Zhang, JCAP \textbf{1412}(12), 018 (2014)

\bibitem{Geng:2015ara}
J.J. Geng, Y.H. Li, J.F. Zhang, X.~Zhang, Eur. Phys. J. \textbf{C75}(8), 356
  (2015)

\bibitem{Zhang:2013zyn}
M.J. Zhang, W.B. Liu, Eur. Phys. J. \textbf{C74}, 2863 (2014)

\bibitem{Geng:2015hen}
J.J. Geng, R.Y. Guo, D.Z. He, J.F. Zhang, X.~Zhang, Front. Phys. (Beijing)
  \textbf{10}, 109501 (2015)

\bibitem{Li:2013oba}
Z.~Li, K.~Liao, P.~Wu, H.~Yu, Z.H. Zhu, Phys. Rev. \textbf{D88}(2), 023003
  (2013)

\bibitem{Zhang:2007zga}
H.b. Zhang, W.~Zhong, Z.H. Zhu, S.~He, Phys. Rev. \textbf{D76}, 123508 (2007)

\bibitem{Zhu:2015pta}
W.T. Zhu, P.X. Wu, H.W. Yu, Chin. Phys. Lett. \textbf{32}(5), 059501 (2015)

\bibitem{Zhang:2014bwt}
M.J. Zhang, J.Z. Qi, W.B. Liu, Int. J. Theor. Phys. \textbf{54}(7), 2456 (2015)

\bibitem{Martins:2016bbi}
C.J.A.P. Martins, M.~Martinelli, E.~Calabrese, M.P.L.P. Ramos, Phys. Rev.
  \textbf{D94}(4), 043001 (2016)

\bibitem{Kim:2014uha}
A.G. Kim, E.V. Linder, J.~Edelstein, D.~Erskine, Astropart. Phys. \textbf{62},
  195 (2015)

\bibitem{Quercellini:2010zr}
C.~Quercellini, L.~Amendola, A.~Balbi, P.~Cabella, M.~Quartin, Phys. Rept.
  \textbf{521}, 95 (2012)

\bibitem{Padmanabhan:2002vv}
T.~Padmanabhan, T.R. Choudhury, Mon. Not. Roy. Astron. Soc. \textbf{344}, 823
  (2003)

\bibitem{Liu:2014vda}
J.~Liu, H.~Wei, Gen. Rel. Grav. \textbf{47}(11), 141 (2015)

\bibitem{Moresco:2016mzx}
M.~Moresco, L.~Pozzetti, A.~Cimatti, R.~Jimenez, C.~Maraston, L.~Verde,
  D.~Thomas, A.~Citro, R.~Tojeiro, D.~Wilkinson, JCAP \textbf{1605}(05), 014
  (2016)

\bibitem{Alam:2016hwk}
S.~Alam, et~al., Mon. Not. Roy. Astron. Soc. \textbf{470}(3), 2617 (2017).
\newblock \doi{10.1093/mnras/stx721}

\bibitem{Risaliti:2015zla}
G.~Risaliti, E.~Lusso, Astrophys. J. \textbf{815}, 33 (2015).
\newblock \doi{10.1088/0004-637X/815/1/33}

\bibitem{CODEX2010a}
CODEX Phase A Science Case, document E-TRE-IOA-573-0001 Issue 1  (2010)

\bibitem{Martinelli:2012vq}
M.~Martinelli, S.~Pandolfi, C.J.A.P. Martins, P.E. Vielzeuf, Phys. Rev.
  \textbf{D86}, 123001 (2012)

\bibitem{Font-Ribera:2013rwa}
A.~Font-Ribera, P.~McDonald, N.~Mostek, B.A. Reid, H.J. Seo, A.~Slosar, JCAP
  \textbf{1405}, 023 (2014)

\bibitem{Fixsen:2009ug}
D.~Fixsen, Astrophys. J. \textbf{707}, 916 (2009)

\bibitem{Hu:1995en}
W.~Hu, N.~Sugiyama, Astrophys. J. \textbf{471}, 542 (1996)

\bibitem{Seo:2007ns}
H.J. Seo, D.J. Eisenstein, Astrophys. J. \textbf{665}, 14 (2007)

\bibitem{Christensen:2001gj}
N.~Christensen, R.~Meyer, L.~Knox, B.~Luey, Class. Quant. Grav. \textbf{18},
  2677 (2001)

\bibitem{Lewis:2002ah}
A.~Lewis, S.~Bridle, Phys. Rev. \textbf{D66}, 103511 (2002)

\bibitem{Trotta:2004qj}
R.~Trotta, {Cosmic microwave background anisotropies: Beyond standard
  parameters}.
\newblock Ph.D. thesis, Geneva U., arXiv:astro-ph/0410115 (2004)

\bibitem{Carroll:1991mt}
S.M. Carroll, W.H. Press, E.L. Turner, Ann. Rev. Astron. Astrophys.
  \textbf{30}, 499 (1992)

\bibitem{Sahni:1999gb}
V.~Sahni, A.A. Starobinsky, Int. J. Mod. Phys. \textbf{D9}, 373 (2000)

\bibitem{Wang2013}
Y.~Wang, S.~Wang, Phys. Rev. \textbf{D88}(4), 043522 (2013)

\bibitem{Knop:2003iy}
R.A. Knop, et~al., Astrophys. J. \textbf{598}, 102 (2003)

\bibitem{Riess:2004nr}
A.G. Riess, et~al., Astrophys. J. \textbf{607}, 665 (2004)

\bibitem{Slepian:2013ug}
Z.~Slepian, J.R. Gott, III, J.~Zinn, Mon. Not. Roy. Astron. Soc.
  \textbf{438}(3), 1948 (2014)

\bibitem{Ade:2015rim}
P.A.R. Ade, et~al., Astron. Astrophys. \textbf{594}, A14 (2016)

\bibitem{Aubourg:2014yra}
E.~Aubourg, et~al., Phys. Rev. \textbf{D92}(12), 123516 (2015)

\bibitem{Chevallier:2000qy}
M.~Chevallier, D.~Polarski, Int. J. Mod. Phys. \textbf{D10}, 213 (2001)

\bibitem{Linder:2002et}
E.V. Linder, Phys. Rev. Lett. \textbf{90}, 091301 (2003)

\bibitem{Sendra:2011pt}
I.~Sendra, R.~Lazkoz, Mon. Not. Roy. Astron. Soc. \textbf{422}, 776 (2012)

\bibitem{Dunkley2005}
J.~Dunkley, M.~Bucher, P.G. Ferreira, K.~Moodley, C.~Skordis, Mon. Not. Roy.
  Astron. Soc. \textbf{356}, 925 (2005)

\bibitem{Coe:2009xf}
D.~Coe, arXiv:0906.4123  (2009)

\bibitem{tension}
C.~Escamilla-Rivera, R.~Lazkoz, V.~Salzano, I.~Sendra, JCAP \textbf{1109}, 003
  (2011)

\bibitem{Salzano2013}
V.~Salzano, S.A. Rodney, I.~Sendra, R.~Lazkoz, A.G. Riess, M.~Postman,
  T.~Broadhurst, D.~Coe, Astron. Astrophys. \textbf{557}, A64 (2013)

\end{thebibliography}
\bibliographystyle{apsrev}

\end{document}